\documentclass[twocolumn,showpacs,preprintnumbers,prc,aps,amsfonts,amsmath,amssymb,nofootinbib]{revtex4}

\usepackage{graphicx}
\usepackage{amsmath}
\usepackage{amsfonts}
\usepackage{amssymb}
\newcommand{\bff}[1]{{\mbox{\boldmath $#1$}}}

\begin{document}
\title{Time-odd mean fields in covariant density functional theory: \\
Non-rotating systems.}
\author{A.\ V.\ Afanasjev and H.\ Abusara}
\affiliation{Department of Physics and Astronomy, Mississippi State 
University, Mississippi State, Mississippi 39762, USA}
\date{\today}

\begin{abstract}
  Time-odd mean fields (nuclear magnetism) are analyzed in the framework 
of covariant density functional theory (CDFT) by performing the 
blocking of the single-particle states with fixed signature. It is shown 
that they always provide additional binding to the binding energies of 
odd-mass nuclei. This additional binding only weakly depends on the RMF 
parametrization reflecting good localization of the properties of 
time-odd mean fields in CDFT. The 
underlying microscopic mechanism is discussed in detail. Time-odd mean fields 
affect odd-even mass differences. However,  our analysis suggests that the 
modifications of the strength of pairing correlations required to compensate 
for their effects are modest.  In contrast, time-odd mean fields have a 
profound effect on the properties of odd-proton nuclei in the vicinity of 
the proton-drip 
line. Their presence can modify the half-lives of  proton-emitters (by many 
orders of magnitude in light nuclei) and considerably affect the possibilities 
of their experimental observation.
\end{abstract}

\maketitle

\section{Introduction}
\label{Sect-intro}

 The development of self-consistent many-body theories aiming at the 
description of low-energy nuclear phenomena provides the necessary 
theoretical tools for an exploration of the nuclear chart into known 
and unknown regions. Theoretical methods (both relativistic and 
non-relativistic) formulated within the framework of density functional 
theory (DFT) and effective field theory (EFT) are the most promising 
tools for the global investigation of the properties of atomic nuclei. The 
DFT and EFT concepts in nuclear structure models have been extensively 
discussed in a number of recent articles \cite{FS.00,LRV.04,F.06,FRS.08}.
The power of the models based on these concepts is essentially 
unchallenged in medium and heavy mass nuclei where 'ab-initio' type
few-body calculations are computationally impossible and the applicability 
of the spherical shell model is restricted to a few regions in the vicinity 
of doubly shell closures.

   The self-consistent mean-field approach to nuclear structure represents 
an approximate implementation of Kohn-Sham density functional theory (DFT)
\cite{HK.64,KS.65,Koh.99,DG.90}, which is successfully employed in the 
treatment of the quantum many-body problem in atomic, molecular and condensed 
matter physics. 
The DFT enables a description of the nuclear many-body problem in terms 
of energy density functionals (EDF), and self-consistent mean-field models approximate 
these functionals, which include all higher-order correlations, with powers 
and gradients of ground-state nucleon densities (see Refs.\ 
\cite{SW.97,BHR.03,VALR.05,F.06,CDK.08} and references therein). EDF functionals 
are universal in the sense that they can be applied to nuclei all over the
periodic table.
 Although they model the effective interaction between nucleons, EDF are not 
necessarily related to any nucleon-nucleon (NN) 
potential. By employing these energy functionals, adjusted to reproduce 
the empirical properties of symmetric and asymmetric nuclear matter, and bulk 
properties of some spherical nuclei, the current generation of self-consistent 
mean-field methods has achieved a high level of accuracy in the description of 
the ground states and the properties of excited states in arbitrarily 
heavy nuclei, exotic nuclei far from $\beta$-stability, and in nuclear systems 
at the nucleon drip-lines (see Refs.\ \cite{BHR.03,VALR.05,SP.07} and 
references therein).

 The self-consistent methods (such as Hartree-Fock (HF) or Hartree-Fock-Bogoliubov 
(HFB)) based on zero range Skyrme forces or finite range Gogny forces are frequently 
used in nuclear structure calculations \cite{RS.80,BHR.03}. These approaches represent 
non-relativistic energy density functionals based on the Schrodinger equation for 
many-body nuclear problem \cite{BHR.03}.

  On the other hand, one can formulate the class of relativistic models 
based on the Dirac formalism, which can generally be defined as {\it
covariant density functionals} (CDF) \cite{VALR.05}. These models, such as quantum  
hadrodynamics (QHD) \cite{SW.86,SW.97}, are based on concepts of non-renormalizable 
effective relativistic field theories and DFT, and they provide a very interesting 
relativistic framework for the studies of nuclear structure phenomena at and far 
from the valley of $\beta-$stability \cite{VALR.05}. Relativistic mean field (RMF) 
models \cite{SW.86} are analogs of the Kohn-Sham formalism of DFT \cite{Koh.99}, 
with local scalar and vector fields appearing in the role of local relativistic 
Kohn-Sham potentials \cite{SW.97,FS.00}. The energy density functional is 
approximated with the powers and gradients of auxiliary meson fields or nucleon densities. 
The EFT building of the energy density functional allows error estimates to be made, 
provides a power counting scheme which separates long- and short-distance dynamics 
and, therefore, removes model dependences from the self-consistent mean field 
approach \cite{FS.00-1}.  In the description of nuclear ground states and the properties 
of excited states the self-consistent mean-field implementations of quantum 
hadrodynamics, the relativistic Hartree-Bogoliubov model (RHB) and the relativistic 
(quasiparticle) random phase approximation (RQRPA) and their subversions, are 
employed \cite{VALR.05}.

 The {\it mean field} is a basic concept of every DFT. One can specify  {\it time-even} 
and {\it time-odd} mean fields \cite{DD.95,AR.00} dependent on the response of these 
fields to the action of time-reversal operator. The properties of time-even mean fields in 
nuclear density  functionals are reasonably well understood and defined \cite{BHR.03,VALR.05}. 
This is due to the facts that (i) many physical observables such as binding energies, radii 
etc. are sensitive only to these fields, and (ii) the model parameters are fitted to such 
physical observables.

 On the other hand, the properties of time-odd mean fields, which appear only 
in nuclear systems with broken time-reversal symmetry, are still poorly understood. 
However, it is already known that these fields are important for proper description 
of rotating nuclei \cite{PWM.85,KR.90,KR.93,DD.95,AR.00}, band terminations \cite{ZSW.05,A.08}, 
magnetic moments \cite{HR.88}, isoscalar monopole vibrations \cite{EBGKV.75},  
electric giant resonances \cite{NKKVR.08}, 
large amplitude collective dynamics \cite{HNMM.06}, 
fussion process \cite{UO.06}, the strengths and energies of Gamow-Teller resonances 
\cite{BDEN.02}, the binding energies of odd-mass nuclei \cite{S.99,DBHM.01,A250} and 
the additivity of angular momentum alignments \cite{MADLN.07}. They also may play a 
role in the $N=Z$ nuclei \cite{DDW.03,S.99} and affect the definition of the strength of 
pairing correlations \cite{RBRM.99,A250}.

  There was a dedicated effort to better understand time-odd mean fields 
in the framework of the Skyrme energy density functional (EDF) theory (see Refs.\ 
\cite{PWM.85,DD.95,ZSW.05,BDEN.02} and references therein). On the contrary, much less 
attention has been paid to these fields in covariant density functional theory (CDFT) 
\cite{HR.88,RBRM.99,AR.00,A250}. This is due to the fact that time-odd mean 
fields are defined through the Lorentz invariance in the CDFT \cite{VALR.05}, and thus 
they do not require additional coupling constants. On the other hand, time-odd mean fields 
are not well defined in non-relativistic density functional theories \cite{DD.95,BDEN.02} 
and, as a consequence, there are a number of open questions related to these fields.
The current 
manuscript aims at better and systematic understanding of time-odd mean fields and
their impact on physical observables in non-rotating nuclei in the framework of the 
RMF realization of the CDFT. The results of the study of these fields in rotating nuclei 
will be presented in a forthcoming article \cite{AA.09-2} which represents a continuation 
of the current investigation.

    The manuscript is organized as follows. The cranked relativistic mean field 
theory and its details related to time-odd mean fields are discussed in Sect.\ 
\ref{theory}. Section \ref{Sect-odd} is devoted to the analysis of the impact of 
time-odd mean fields on binding energies of odd-mass nuclei. The mass and particle
number dependences of this impact and their connections with odd-even mass 
staggerings are also considered. The microscopic mechanism of 
additional binding in odd-mass nuclei induced by time-odd mean fields is analyzed 
in Sect.\ \ref{Mech-bind-to}. The impact of time-odd mean fields on the properties of 
proton-unstable nuclei is studied in Sect.\ \ref{Sect-proton}. Section \ref{Sect-odd-odd} 
considers how time-odd mean fields modify the properties of odd-odd nuclei. 
 Finally, Sect.\ \ref{Sect-concl} contains the main conclusions of our work.

\section{Theoretical formalism}
\label{theory}


 The results presented in the current manuscript have been obtained using
Cranked Relativistic Mean Field (CRMF) theory \cite{KR.89,KR.93,AKR.96}. 
This theory has been successfully employed for the description of rotating 
nuclei (see Ref.\ \cite{AA.08} and references therein) in which time-odd 
mean fields play an important role, but it is also able to describe the nuclear 
systems with broken time-reversal symmetry in intrinsic frame at no rotation.
In this theory the pairing correlations are neglected which allows to better 
isolate the effects induced by time-odd mean fields. 
The CRMF computer code is formulated in the signature basis. As a result, the 
breaking of Kramer's degeneracy of the single-particle states is taken 
into account in a fully self-consistent way. This is important for an accurate 
description of time-odd mean fields in fermionic channel (see Sect.\ 
\ref{Mech-bind-to}). The most important features of the CRMF formalism related to 
time-odd mean fields are outlined below (for more details see Refs.\ 
\cite{KR.89,AKR.96}) for the case of no rotation (rotational frequency 
$\Omega_x=0)$.

 In the Hartree approximation, the stationary Dirac equation for the
nucleons in the intrinsic frame is given by
\begin{eqnarray}
\hat{h}_{D} \psi_i=\varepsilon_i \psi_i
\end{eqnarray}
where $\hat{h}_{D}$ is the Dirac Hamiltonian for the nucleon with mass $m$
\begin{eqnarray}
\hat{h}_{D}={\bff{\alpha}}(-i{\bff{\nabla}}-{\bff{V}}({\bff r}))~+~V_{0}({\bff r})~
+~\beta (m+S({\bff r})).
\end{eqnarray}
It contains the average fields determined by the mesons, i.e. the attractive 
scalar field $S({\bf r})$
\begin{eqnarray}
S({\bff r})=g_{\sigma} \sigma ({\bff r}),
\end{eqnarray}
and the repulsive time-like component of the vector field $V_{0}({\bff r})$
\begin{eqnarray}
V_0({\bff r}) = g_{\omega} \omega_0({\bff r}) + g_{\rho} \tau_3 \rho_0 ({\bff r}) + e 
\frac{1-\tau_3}{2} A_0 ({\bff r}).
\end{eqnarray}  
A magnetic potential ${\bff V} ({\bff r})$ 
\begin{eqnarray}
\bff{V}({\bff r})=g_{\omega }{\bff{\omega}}({\bff r})+ g_{\rho}\tau_3{\bff{\rho}}
({\bff r})+e\frac{1-\tau _{3}}{2}\bff{A}({\bf r}),
\label{magnetic}
\end{eqnarray}
originates from the space-like components of the vector mesons. Note that in these
equations, the four-vector components of the vector fields $\omega^{\mu}$, 
$\rho^{\mu}$, and $A^{\mu}$ are separated into the time-like ($\omega_0$, $\rho_0$
and $A_0$) and space-like [${\bff \omega}=(\omega^x, \omega^y, \omega^z)$, 
${\bff \rho}=(\rho^x, \rho^y, \rho^z)$, and ${\bff A}=(A^x, A^y, A^z)$] components. 
In the Dirac equation the magnetic potential has the structure of a magnetic field. 
Therefore the effect produced by it is called \textit{nuclear magnetism } (NM) 
\cite{KR.89}. 

  The corresponding meson fields and the electromagnetic potential are determined by 
the Klein-Gordon equations
\begin{eqnarray}
\left\{ -\Delta
+m_{\sigma}^{2}\right
\} ~\sigma({\bff r}) & = & -g_{\sigma} [\rho^n_s({\bff r}) + \rho^p_s ({\bff r})]
\nonumber \\
& & -g_{2}\sigma^{2}({\bff r})-g_{3}\sigma^{3}({\bff r}), \label{KGsigma} \\
\left\{ -\Delta
+m_{\omega}^{2}\right\}
\omega_{0}({\bff r}) & = &g_{\omega} [\rho_v^n({\bff r}) + \rho_v^p({\bff r})],  
 \label{KGomega0} \\
\left\{ -\Delta
+m_{\omega }^{2}\right\}
~{\bff{\omega}}({\bff r}) & = &g_{\omega} [{\bff j}^n({\bff r}) + {\bff j}^p({\bff r})]
\label{KGomegav} \\
\left\{ -\Delta
+m_{\rho}^{2}\right\}
\rho_{0}({\bff r}) & = & g_{\rho} [\rho_v^n({\bff r}) - \rho_v^p({\bff r})],  
 \label{KGRho0} \\
\left\{ -\Delta
+m_{\rho}^{2}\right\}
~{\bff{\rho}}({\bff r}) & = &g_{\rho} [{\bff j}^n({\bff r}) - {\bff j}^p({\bff r})],
\label{KGRhov} \\
-\Delta A_0({\bff r})=e\rho^p_v({\bff r}) , & &  -\Delta {\bff A}({\bff r})=e{\bff j}^p({\bff r}),
\end{eqnarray}
with source terms involving the various nucleonic densities and currents
\begin{eqnarray}
\rho_s^{n,p}({\bff r}) &=& \sum_{i=1}^{N,Z} (\psi_i({\bff r}))^{\dagger}
\hat{\beta} \psi_i({\bff r}),  \\
\rho_v^{n,p}({\bff r}) &=& \sum_{i=1}^{N,Z} (\psi_i({\bff r}))^{\dagger}
\psi_i({\bff r}),  \\
{\bff j}^{n,p}({\bff r}) &=& \sum_{i=1}^{N,Z} (\psi_i({\bff r}))^{\dagger}
\hat{{\bff \alpha}} \psi_i({\bff r}) \label{current}
\end{eqnarray}
where the labels $n$ and $p$ are used for neutrons and protons, respectively. 
In the equations above, the sums run over the occupied positive-energy shell 
model states only ({\it no-sea approximation}) \cite{SW.86,NL1}. Note that the 
spatial components of the vector potential ${\bff A}({\bff r})$ are neglected 
in the calculations since the coupling constant of the electromagnetic 
interaction is small compared with the coupling constants of the meson fields.

\begin{figure}
\vspace{0.5cm}
\centering
\includegraphics[width=8.0cm]{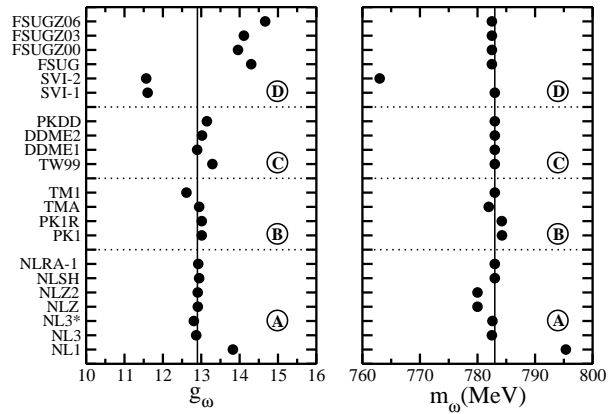}
\caption{ The $m_{\omega}$ and $g_{\omega}$ parameters of different modern 
parametrizations of the RMF Lagrangian. They are combined into four groups dependent 
on how self- and mixed-couplings are introduced.  Group A represents the parametrizations 
which include non-linear self-couplings only for the $\sigma$-meson.  Group B contains
the parametrizations which include self-couplings for the $\sigma$- and $\omega$-mesons
(and $\rho-$mesons in the case of PK1R). Group C represents the parametrizations which 
include density-dependent meson-nucleon couplings for the $\sigma$-, $\omega$-, and 
$\rho$-mesons. The other parametrizations are included into the group D. 
The SVI and SVI-2 parametrizations neglect non-linear self-couplings for the $\sigma$-meson 
but include isoscalar-isovector couplings. The other parametrizations of the group D include mixed 
interaction terms (such as isoscalar-isovector couplings \cite{FSUGold}) in addition to 
non-linear self-couplings for the $\sigma$-meson. The parameters are taken from Refs.\ 
\cite{NL1} (NL1), \cite{NL3} (NL3), 
\cite{NL3*} (NL3*), \cite{NLZ} (NLZ), \cite{NLZ2} (NLZ2), \cite{NLSH} (NLSH),
\cite{NLRA1} (NLRA-1), \cite{PK1} (PK1,PK1R), \cite{TMA} (TMA), \cite{TM1} (TM1), 
\cite{SVI1} (SVI1,SVI-2), \cite{FSUGold} (FSUGold), \cite{FSUG00} (FSUGZ00,FSUGZ03,FSUGZ06), 
\cite{TW99} (TW99), \cite{NVFR.02} (DDME1), \cite{LNVR.05} (DDME2), \cite{PK1} (PKDD).
Note that we omitted mass-dependent terms for $g_{\omega}$ in the TMA parametrization which 
is a good approximation for heavy nuclei since $g_{\omega}=12.842+3.191A^{-0.4}$ \cite{TMA}.}
\label{Paramet}
\end{figure}

   The magnetic potential ${\bff V} ({\bff r})$ in the Dirac equation as well as the currents 
${\bff j}^{n,p}({\bff r})$ in the Klein-Gordon equations do not appear in the RMF equations for 
time-reversal systems \cite{SW.86}. Similar to the nonrelativistic case, their presence leads 
to the appearance of time-odd mean fields. Thus, we will use the terms {\it nuclear magnetism} and 
{\it time-odd mean fields} interchangeably throughout this manuscript.
The  magnetic potential is the contribution to the mean field that breaks 
time-reversal symmetry in the intrinsic frame and induces non-vanishing 
currents $\bff{j}^{n,p}$ (Eq.\ (\ref{current})) in the Klein-Gordon equations 
(Eqs.\ (\ref{KGomegav}), (\ref{KGRhov})), which are related to the space-like 
components of the vector mesons. In turn, the space-like components of
the vector ${\bff \omega}$ and ${\bff \rho}$ fields form the magnetic 
potential (\ref{magnetic}) in the Dirac equation. Note that the current
$\bff{j}^{n,p}({\bff r})$ change the sign upon the action of time-reversal 
operator  \cite{G.90}. Together with densities it forms covariant 
four-vector $j^{\mu}=\{\rho, \bff{j}\}$. As a consequence, these two quantities 
($\rho$ and $\bff{j}$) do not transform independently under Lorentz transformation.
This explains why  the structure of the Klein-Gordon equations for time-like 
and space-like components of vector mesons is the same (compare, for example, 
Eqs.\ (\ref{KGomega0}) and (\ref{KGomegav}) for $\omega$-meson) and why the same coupling constant 
stands in front of the densities and currents on the right hand side of these 
equations.

  The spatial components of the vector $\omega$ and $\rho$ mesons lead to the
interactions between possible currents. For the $\omega$-meson this interaction 
is attractive for all combinations ($pp$, $nn$ and $pn$-currents), and for the 
$\rho$-meson it is attractive for $pp$ and $nn$-currents but repulsive for $pn$-currents. 
Within mean field theory such currents occur only in the situations of broken 
time-reversal symmetry.

  Note that time-odd mean fields related to NM are defined through the Lorentz 
invariance \cite{VALR.05} and thus they do not require additional coupling constants: 
the coupling constants of time-even mean fields are used also for time-odd mean 
fields.

   The currents are isoscalar and isovector in nature for the $\omega$ and $\rho$ 
mesons (Eqs.\ (\ref{KGomegav}, \ref{KGRhov})), respectively. As a consequence, the 
contribution of the $\rho$-meson to magnetic potential and total energy is 
marginal in the majority of the cases even at the neutron-drip line (see Sect.\ 
\ref{Pol-odd} for details).
Thus, time-odd mean fields in the RMF framework depend predominantly 
on the spatial components of the $\omega$ meson. Neglecting the contribution of the
$\rho$ meson, one can see that only two parameters, namely, the mass $m_{\omega}$  and 
coupling constant $g_{\omega}$ of the $\omega$ meson define the properties of time-odd 
mean fields (Eqs.\ (\ref{magnetic}), (\ref{KGomegav}), and (\ref{KGRhov})). Fig.\ 
\ref{Paramet} clearly indicates that these parameters are well localized in the 
parameter space for the parametrizations of the RMF Lagrangian  in the
groups A, B, and C (see caption of Fig.\ \ref{Paramet} for group definitions). 
This suggests that the parameter dependence of the impact of time-odd  mean fields 
on the physical observables should be quite weak for these types of the 
parametrizations. Indeed, the analysis of terminating states in Ref.\ \cite{A.08} 
showed that time-odd mean fields are defined with an accuracy around 15\%  for 
the parametrizations of the RMF Lagrangian containing only non-linear self-couplings 
for the $\sigma$-meson (group A). 

 On the other hand, Fig.\ \ref{Paramet} suggests that the time-odd mean fields maybe
less accurately defined in the parametrizations of group D which include mixed 
interaction terms such as isoscalar-isovector couplings. However, it is premature
to make such conclusion because these parametrizations have not been tested extensively 
even on nuclear structure data sensitive to time-even mean fields. This is contrary to 
the parametrizations of groups A, B, and C which have passed successfully such test.

 The investigation of all these parametrizations is definitely beyond the scope
of this study. Thus, the present investigation has been focused on the study of 
time-odd mean fields in the CDFT with the parametrizations of the RMF Lagrangian
including only non-linear self-couplings of the $\sigma$-meson (the group A of 
the parametrizations). The results of the study of time-odd mean fields in the 
groups B, C, and D of the parametrizations of meson-coupling models as well as
within the point-coupling models will be presented in a forthcoming manuscript 
\cite{AA.09-2}.

 The total energy of the system is given in Refs.\ \cite{KR.89,AKR.96}. In order to 
facilitate the discussion we split it into different terms as \footnote{We follow Refs.\ 
\cite{RGL.97,R-priv-09} in the selection of the signs of the energy terms.}
\begin{eqnarray}
E_{tot} & = & E_{part} + E_{cm} - E_{\sigma} - E_{\sigma NL} - E_{\omega}^{TL} - 
E^{TL}_{\rho} \nonumber \\
       &   &  - E_{\omega}^{SL} - E^{SL}_{\rho} - E_{Coul},
\label{Etot}
\end{eqnarray}
where $E_{part}$ and $E_{cm}$ represent the contributions from fermionic
degrees of freedom,  while the other terms are related to mesonic (bosonic) 
degrees of freedom.
In Eq.\ (\ref{Etot})
\begin{eqnarray}
E_{part}=\sum_i^A\varepsilon_i ,
\label{Epart}
\end{eqnarray}
is the energy of the particles moving in the field created by the mesons ($\varepsilon_i$ 
is the energy of $i$-th particle and the sum runs over all occupied proton and neutron 
states)
\begin{eqnarray}
E_{\sigma}=\frac{1}{2}\,\, g_{\sigma} \int d^3r\,\, \sigma({\bff r}) \left[\rho_s^p({\bff r})+\rho_s^n({\bff r})\right] ,
\end{eqnarray}
is the linear contribution to the energy of isoscalar-scalar $\sigma$-field
\begin{eqnarray}
E_{\sigma NL} = \frac{1}{2} \int d^3r \left[ \frac{1}{3}\,g_2\, \sigma^3({\bff r}) + \frac{1}{2}\, g_3\, \sigma^4({\bff r}) 
\right] ,
\end{eqnarray}
is the non-linear contribution to the energy of isoscalar-scalar $\sigma$-field
\begin{eqnarray}
E^{TL}_{\omega} = \frac{1}{2}\,\, g_\omega \int d^3r\,  \omega_0({\bff r}) \left[ \rho_v^p({\bff r})+\rho_v^n({\bff r}) \right] ,
\end{eqnarray}
is the energy of the time-like component of isoscalar-vector $\omega$-field
\begin{eqnarray}
E^{TL}_{\rho} = \frac{1}{2} \,\,g_\rho \int d^3r  \rho_0({\bff r}) \left[\rho_v^n({\bff r})-\rho_v^p({\bff r})\right] ,
\end{eqnarray}
is the energy of the time-like component of isovector-vector $\rho$-field
\begin{eqnarray}
E^{SL}_{\omega} = - \frac{1}{2}\,\,  g_\omega \int d^3r\, {\bff \omega}({\bff r}) \left[ {\bff j}^p({\bff r})+{\bff j}^n({\bff r}) 
\right] ,
\label{E-s-omega}
\end{eqnarray}
is the energy of the space-like component of isoscalar-vector $\omega$-field
\begin{eqnarray}
E^{SL}_{\rho} = - \frac{1}{2}\,\, g_\rho \int d^3r\,  {\bff \rho}({\bff r}) \left[ {\bff j}^n({\bff r})-{\bff j}^p({\bff r}) 
\right] ,
\label{rho-space}
\end{eqnarray}
is the energy of the space-like component of isovector-vector $\rho$-field
\begin{eqnarray}
E_{Coul} = \frac{1}{2}\,\,  e \int d^3r  A_0({\bff r}) \rho^p_v({\bff r}) ,
\end{eqnarray}
is the Coulomb energy
\begin{eqnarray}
E_{cm}=-\frac{3}{4}\hbar \omega_0=-\frac{3}{4}\,\, 41 A^{-1/3}\,\, {\rm MeV} ,
\end{eqnarray}
is the correction for the spurious center-of-mass motion approximated by its value in
a non-relativistic harmonic oscillator potential.

  The total energy of the system can alternatively be written as (similar 
to Refs.\ \cite{RGL.97,R-priv-09})
\begin{eqnarray}
E_{tot} = E_{kin} + E_{int} + E_{cm}
\end{eqnarray}
where the kinetic energy $E_{kin}$ is given by
\begin{eqnarray}
E_{kin}=E_{part} - 2\, (E_{\sigma} + E_{\omega}^{TL} + E_{\rho}^{TL} + E_{Coul})
\label{Ekin-f}
\end{eqnarray}
and the interaction energy between the nucleons $E_{int}$ by
\begin{eqnarray}
E_{int} &=& E_{\sigma} + E_{\omega}^{TL} + E_{\rho}^{TL} + E_{Coul} \nonumber \\
        && - E_{\sigma NL} - E_{\omega}^{SL} - E^{SL}_{\rho}.
\end{eqnarray}
 However, this representation of total energy has a disadvantage as
compared with Eq.\ (\ref{Etot}) since it does not provide a direct access 
to the particle energy $E_{part}$. The latter plays an important role in the 
understanding of the breaking of Kramer's degeneracy of time-reversal 
orbitals in the presence of time-odd mean fields (see Sect.\ \ref{E-split} 
for details). Thus, further discussion of total energy will be based  
mostly on Eq.\ (\ref{Etot}). However, we will also provide the results of
calculations for kinetic energy $E_{kin}$. 

   The CRMF equations are solved in the basis of an anisotropic three-dimensional 
harmonic oscillator in Cartesian coordinates characterized by the deformation 
parameters $\beta_0=0.3$ ($\beta_0=0.4$ in the case of superdeformed states) and 
$\gamma=0^{\circ}$ as well as the oscillator frequency $\hbar \omega_0= 41 A^{-1/3}$ MeV. 
The truncation of basis is performed in such a way that all states belonging to 
the shells up to fermionic  $N_F$=12 and bosonic $N_B$=16 are taken into account
in the calculations of light and medium-mass nuclei. The fermionic basis is increased
up to $N_F$=14 in the calculations of actinides. Numerical analysis indicates that this 
truncation scheme provides sufficient numerical accuracy for the physical quantities
of interest.

Single-particle orbitals are labeled by $[Nn_z\Lambda]\Omega^{sign}$. 
$[Nn_z\Lambda]\Omega$ are the asymptotic quantum numbers (Nilsson quantum numbers) of 
the dominant component of the wave function. The superscripts {\it sign} to the orbital 
labels are used sometimes to indicate the sign of the signature $r$ for that orbital 
$(r=\pm i)$. The majority of the calculations are performed with the NL3 parametrization 
\cite{NL3} of the RMF Lagrangian.
  
 Many-particle configurations (further nuclear configurations or configurations) 
are specified by the occupation of available single-particle orbitals. In the calculations 
without pairing, the occupation numbers $n$ are integer ($n=0$ or 1). In odd nuclei, all 
single-particle states with exception of one are pairwise occupied. We 
will call this occupied single-particle state of fixed signature for which its 
time-reversal (signature) counterpart state is empty as {\it blocked state} in 
order to simplify the discussion. The total signature and the parity of the 
configuration are the same as the ones of the blocked state. In the CRMF code, it 
is possible to specify the occupation of either $r=+i$ or $r=-i$ 
signature of the single-particle state. The specification of nuclear configuration 
by means of listing all occupied single-particle states is unpractical. Thus, we 
label the nuclear configuration in odd mass nuclei by the 
Nilsson label and the signature of the blocked state. Note that many physical 
observables, such as additional binding due to NM, do not depend on the signature 
of the blocked state in odd-mass nuclei. In these cases, we will omit the signature 
from the configuration label. In odd-odd nuclei, the Nilsson labels of the 
blocked proton and neutron states and their signatures are used for configuration 
labelling. Note that the labelling by means of Nilsson labels is performed only 
when the calculated shape of nuclear configuration is prolate or near-prolate.

 In order to investigate the impact of NM (time-odd mean fields) on physical 
observables, the CRMF calculations are performed in three calculational schemes
for the fixed configurations:
\begin{itemize}
\item
fully self-consistent calculations with NM included (further denoted as NM calculations
\footnote{This method is equivalent to the HF method of Ref.\ \cite{DBHM.01}.}) 
which take into account space-like components of the vector mesons (Eqs.\ (\ref{KGomegav}), 
(\ref{KGRhov}) and (\ref{magnetic})), currents (Eqs.\ (\ref{KGomegav}), (\ref{KGRhov}), 
and (\ref{current})), and magnetic potential {\bff V}({\bff r}) (Eq.\ (\ref{magnetic}));

\item
fully self-consistent calculations without NM (further denoted as WNM calculations
\footnote{The difference between the WNM method of the present manuscript and 
the HFE method of Ref.\ \cite{DBHM.01} is only technical: it is related to the 
treatment of the occupation of the pair of energy degenerate states occupied by 
odd particle. The opposite signature states of this pair are occupied by odd particle 
with probability 0.5 (in the filling approximation) in the HFE method (see Sect. IIC of 
Ref.\ \cite{DBHM.01}). On the contrary, one signature state of this pair
is occupied with probability 1 while other is empty in the WNM method. However,
time-reversal invariance is conserved in both approaches. This means that the 
Kramers degeneracy of the single-particle levels is not violated and time-odd mean 
fields are not introduced into the system. As a consequence, if employed within 
the same framework these two methods
lead to the same results in the calculations without pairing since polarization 
and other  effects induced by odd particle do not depend on the signature of the occupied 
state.}) 
which omit space-like components of the vector mesons (Eqs.\ (\ref{KGomegav}), 
(\ref{KGRhov}) and (\ref{magnetic})), currents (Eqs.\ (\ref{KGomegav}), (\ref{KGRhov}), 
and (\ref{current})), and magnetic potential {\bff V}({\bff r}) (Eq.\ (\ref{magnetic})).
Note that the results of the NM and WNM calculations are always compared for the same 
nuclear configuration;

\item
perturbative calculations (the physical quantities of interest are indicated 
by superscript {\it pert}). Fully self-consistent calculations with NM 
provide a starting point. Using their fields as input fields, only one 
iteration is performed in the calculations without NM: this provides
perturbative results. Time-even mean fields are the same in both (fully 
self-consistent and perturbative) calculations. Then, the impact of time-odd 
mean fields on calculated quantities (for example, different terms in the 
total energy (see Eq.\ \ref{Etot})) is defined as the difference between 
the values of this quantity obtained in these two calculations. In this
way, the pure effects of time-odd mean fields in fermionic and mesonic 
channels of the model are isolated because no polarization effects are 
introduced into time-even mean fields.

\end{itemize}

These are the ways in which the effects of time-odd mean fields can be 
studied, and as such they are frequently used in the DFT studies, both 
in relativistic and non-relativistic frameworks 
\cite{DD.95,S.99,YM.00,DBHM.01,HR.88,KR.93,AR.00,A.08}. One should, however, 
keep in mind that if time-odd fields are neglected, the local Lorentz  
invariance (Galilean invariance in non-relativistic framework 
\cite{DD.95,CDK.08}) is violated. The inclusion of time-odd mean 
fields restores the Lorentz invariance.

\section{Binding energies in odd mass nuclei}
\label{Sect-odd}

  The time-reversal invariance is conserved in the
ground states of even-even nuclei. The nucleon states are then pairwise
degenerated, and the contribution of the state to the currents cancels with
the contribution of its time-reversed partner. Time-odd mean fields reveal 
themselves in odd- and odd-odd  mass nuclei and in two-(multi-)quasiparticle 
states of even-even nuclei. This is because an unpaired (odd) nucleon breaks 
the time-reversal invariance in intrinsic frame and produces the contribution 
to the currents and spin. In this case, the Kramer's degeneracy of time-reversal 
partner orbitals is also broken.

\begin{figure}
\centering
\includegraphics[width=8.0cm]{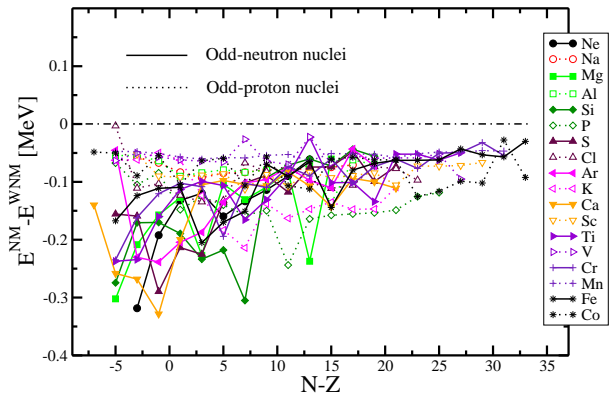}
\caption{(Color online) The impact of NM on binding energies of light 
odd-mass nuclei. The calculations have been performed with the NL3 
parametrization of the RMF Lagrangian. They cover the nuclei from the 
proton-drip line up to the neutron-drip line.}
\label{light-A}
\end{figure}

   The modifications of the binding energies and quasiparticle spectra 
are the most important issues when considering time-odd mean fields in non-rotating 
systems. The binding energies are important in nuclear astrophysics applications 
\cite{LW.01}, and their modifications due to time-odd mean fields may have considerable 
consequences for the $r$- and $rp$-process abundances. 
Thus, it is important to understand the influence of time-odd mean fields on 
binding energies of odd- and odd-odd mass nuclei, especially in the context 
of mass table fits \cite{S.05}. With the current focus on the spectroscopic quality DFT 
\cite{ZDSW.08}, the knowledge on how time-odd mean fields influence the relative 
energies of different (quasi)particle states in model calculations is also needed.

\begin{figure}
\vspace{0.5cm}
\centering
\includegraphics[width=8.0cm]{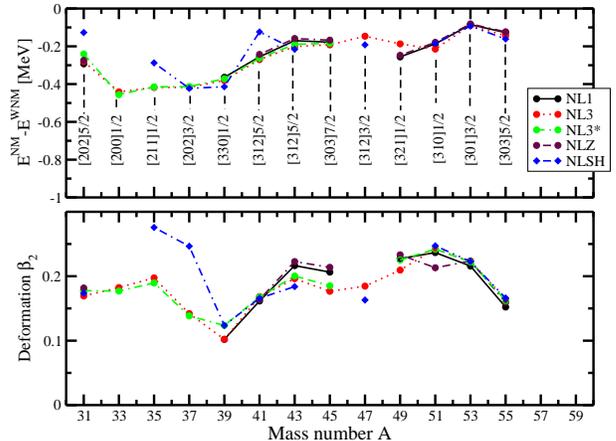}
\caption{(Color online) The same as in Fig.\ \ref{light-A} but for the results obtained 
with the indicated parametrizations of the RMF Lagrangian in the Ar isotopes. The structure 
of the ground states is shown by the Nilsson labels. Only when the structure 
of the ground state is the same as in the NL3 parametrization, the results obtained 
with other parametrizations are shown.}
\label{light-par}
\end{figure}

   While there was a considerable interest in the study of time-odd mean fields in 
odd- and odd-odd mass nuclei at no rotation within the Skyrme EDF \cite{S.99,DBHM.01}, 
relatively little is known about their role in the framework of the CDFT. So far, 
the impact of time-odd mean fields on binding  energies has been studied in the CDFT 
framework only in odd-mass nuclei around doubly magic spherical nuclei in Ref.\ 
\cite{RBRMG.98}, and in few deformed nuclei around $^{32}$S \cite{Pingst-A30-60} 
and $^{254}$No \cite{A250}.

\subsection{Binding energies in light nuclei}
\label{Bind-light}

   The impact of NM on the binding energies of light odd-mass nuclei is 
shown in Fig.\ \ref{light-A}. One can see that in all cases the presence 
of NM leads to additional binding the magnitude of which is nucleus 
and state dependent. The absolute value of this additional binding is 
typically below 200 keV and only reaches 300 keV in some lower mass
nuclei. On the average, the magnitude of additional binding due to NM is 
inversely correlated with the mass of the nucleus; it is the largest in the 
lightest nuclei and the smallest in the heaviest nuclei. For each isotope chain, it is 
the largest in the vicinity of the proton-drip line and the smallest in the vicinity 
of the neutron-drip line. The polarization effects induced by NM and the 
energy splitting between blocked state and its unoccupied signature partner 
induced by NM decrease with the increase of mass (compare Tables \ref{NM-odd} 
and \ref{NM-odd-3} below). This explains the observed trends in additional 
binding due to NM.

  Fig.\ \ref{light-par} shows that additional binding due to NM only weakly 
depends on the RMF parametrization; this is also seen in the analysis of 
terminating states in Ref.\ \cite{A.08}. In both cases, the largest 
deviation from the NL3 results is observed in the case of the NLSH 
parametrization.

  It is interesting to compare these results with the ones obtained in the
Skyrme EDF (see Fig.\ 4 in Ref.\ \cite{S.99}). The modifications of total
binding energy due to time-odd mean fields are given by the $E^{to}$ quantity
in Ref.\ \cite{S.99}, which is an analog of the $E^{NM}-E^{WNM}$ quantity. The 
general dependence of both quantities on $N-Z$ is similar in odd-mass nuclei 
apart from a few cases such as $^{43}$Ti and $^{43}$Sc in SLy4 Skyrme EDF (Fig.\ 4 
in Ref.\ \cite{S.99}). Neither RMF nor Skyrme EDF calculations in odd-mass
nuclei indicate the enhancement of time-odd mean fields in the vicinity of the 
$N=Z$ line. This is contrary to Ref.\ \cite{S.99} which suggested that the 
effects of time-odd mean fields are enhanced at the $N=Z$ line.  The absolute 
values of $E^{to}$ and  $E^{NM}-E^{WNM}$ quantities are similar being below 
300 keV in the majority of the cases.  The principal difference between the RMF 
and the Skyrme EDF lies in the fact that time-odd mean fields are always attractive  
and show very small dependence on the parametrization in the RMF calculations 
(this is also supported by the analysis of terminating states, 
see Ref.\ \cite{A.08}), while they can be both attractive (SLy4 force) or repulsive 
(SIII force) and show considerable dependence on the parametrization in Skyrme EDF 
(Ref.\ \cite{S.99}). 
   
\begin{figure}
\centering
\includegraphics[width=8.0cm]{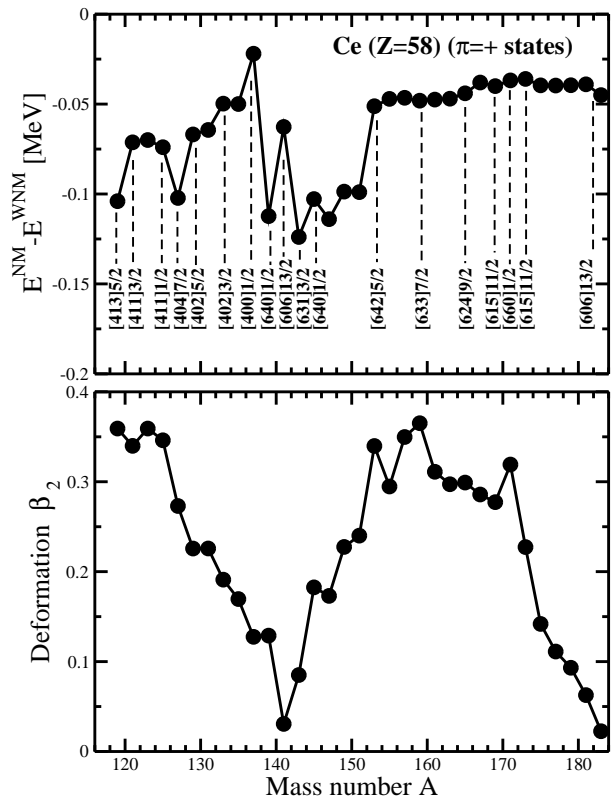}
\caption{The impact of NM on the binding energies of positive parity configurations 
in odd mass Ce ($Z=58$) nuclei. The upper panel shows additional binding $E^{NM}-E^{WNM}$ 
due to NM and its configuration dependence. The configurations are labelled by the
Nilsson labels of the blocked states; the configurations at and to the right of the Nilsson 
label up to the next Nilsson label have the same blocked state. The bottom panel shows the 
corresponding deformations of the configurations. The calculations have been performed with 
the NL3 parametrization of the RMF Lagrangian. They cover the nuclei from the proton-drip 
line up to the neutron-drip line.}
\label{Ce-pos}
\end{figure}

\begin{figure}
\centering
\includegraphics[width=8.0cm]{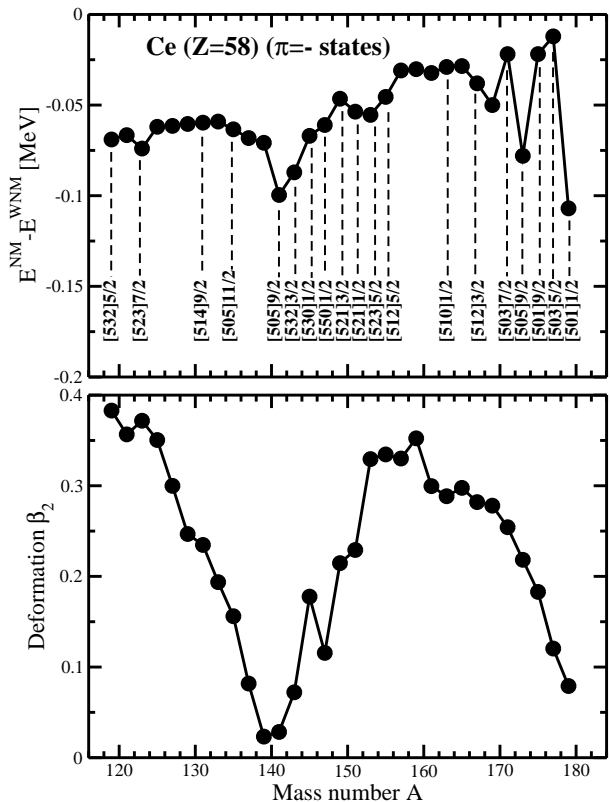}
\caption{The same as in Fig.\ \ref{Ce-pos} but for negative parity 
configurations.}
\label{Ce-neg}
\end{figure}

\subsection{Binding energies in the Ce $(Z=58)$ isotopes}
\label{Ce-isotopes}

  The role of time-odd mean fields is studied here in medium mass Ce isotopes in 
order to facilitate the comparison with the results obtained within the Skyrme EDF 
with the SLy4 force in Ref.\ \cite{DBHM.01}. This reference represents the most detailed 
study of time-odd mean fields in odd-mass nuclei within the Skyrme EDF. We consider 
the lowest configurations of positive and negative parities, while Ref.\ \cite{DBHM.01} 
studies only the lowest configurations in each nucleus. 

  Figs.\ \ref{Ce-pos} and \ref{Ce-neg} show the additional binding due to NM. 
The comparison with the Skyrme EDF results of Ref.\ \cite{DBHM.01} reveals a 
number of important differences.  First, similar to the results in light nuclei 
(Sect.\ \ref{Bind-light}) and in actinide region (Sect.\ VI H in Ref.\ \cite{A250}), 
time-odd mean fields are attractive in the RMF calculations for the Ce isotopes. 
On the contrary, they are repulsive in the SLy4 parametrization of the Skyrme EDF 
\cite{DBHM.01}. Note that the SLy4 force produces attractive 
time-odd mean fields in light nuclei (Ref.\ \cite{S.99}). This mass dependence of 
the effects of time-odd mean fields in the Skyrme EDF may be due to the competition 
between isovector and isoscalar effects \cite{DBHM.01}. The average absolute magnitude 
of the change of binding due to time-odd mean fields in the RMF calculations is only 
half of the one seen in the Skyrme calculations with the SLy4 parametrization. It was 
also checked on some examples that additional binding due to NM only weakly depends 
on the parametrization of the RMF Lagrangian. 
 
  Second, the results of the calculations do not reveal a strong dependence of 
additional binding due to NM on deformation. For example, the deformation of the 
$\nu [615]11/2$ configuration in the $^{173-181}$Ce chain changes drastically from 
$\beta_2\sim 0.23$ down to $\beta_2\sim 0.06$ (Fig.\ \ref{Ce-pos}, bottom panel), 
but the additional binding due to NM remains almost the same (Fig.\ \ref{Ce-pos}, 
top panel). The $\nu [523]7/2$ and $\nu [505]11/2$ configurations are another 
examples of this feature (Fig.\ \ref{Ce-neg}).

  Third, the binding energy modifications due time-odd mean fields are completely 
different in the RMF and Skyrme EDF calculations. In the Skyrme EDF calculations, 
the magnitude of these binding energy modifications is related with three properties 
of the blocked orbital. In decreasing order of importance they are \cite{DBHM.01}: 
a small $\Omega$ 
quantum number, a down-sloping behavior of the energy of the single-particle state 
with mass number $A$, and a large total angular momentum $j$ for the spherical shell 
from which the single-particle state originates.  For example,  the binding energy 
modifications due to time-odd mean fields will be larger for the configuration based 
on blocked single-particle state with small $\Omega$ than for the configuration
with large $\Omega$  of the blocked state if both blocked states belong to the same 
$j$-shell. On the contrary, the RMF calculations do not reveal this type of correlations
between additional binding due to NM and the structure of the blocked state.
Indeed, the configurations which have the largest changes 
in binding energies due to NM ($|E^{NM}-E^{WNM}| \geq 0.1$ MeV) are [413]5/2, 
[404]7/2, [640]1/2, [631]3/2, [505]9/2 and [501]1/2.

\begin{figure*}
\vspace{0.0cm}
\centering
\includegraphics[width=18.0cm]{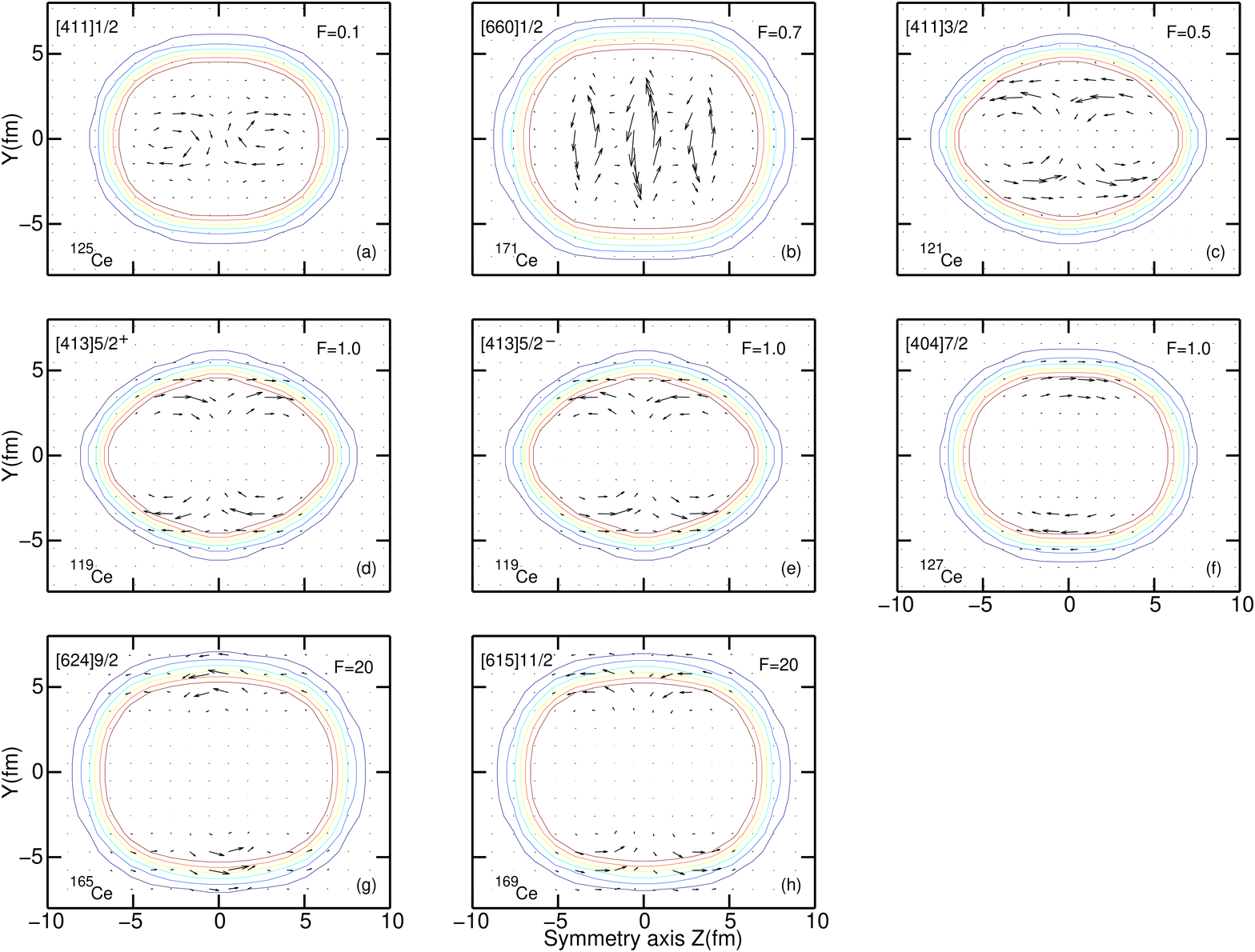}
\vspace{0.0cm}
\caption{(Color online) Total neutron current distributions  {\bff j}$^n$({\bff r}) in 
intrinsic frame in the $y-z$ plane at $x=0.48$ fm for several configurations with 
signature $(r=+i)$ in the Ce nuclei. The only exceptions are the $[413]5/2^{\pm}$
configurations in $^{119}$Ce for which both signatures are shown in panels (d,e).
The Nilsson quantum numbers $[Nn_z\Lambda]\Omega$ indicate the blocked state. 
The currents in panels (d) 
and (e) are plotted at arbitrary units for better visualization. The currents in other 
panels are normalized to the currents in panels (d) and (e) by using factor F. This 
factor is chosen  in such way that the current distribution for every nucleus is 
clearly seen. The shape and size of the nucleus are indicated by density lines
which are plotted in the range $0.01-0.06$ fm$^{-3}$ in step of 0.01 fm$^{-3}$. 
The panels are arranged in such way that the $\Omega$ value of the Nilsson label 
of the blocked state increases on going from panel (a) to panel (h).}
\label{Curr-distr}
\end{figure*}

\begin{figure*}
\vspace{0.0cm}
\centering
\includegraphics[width=14.0cm]{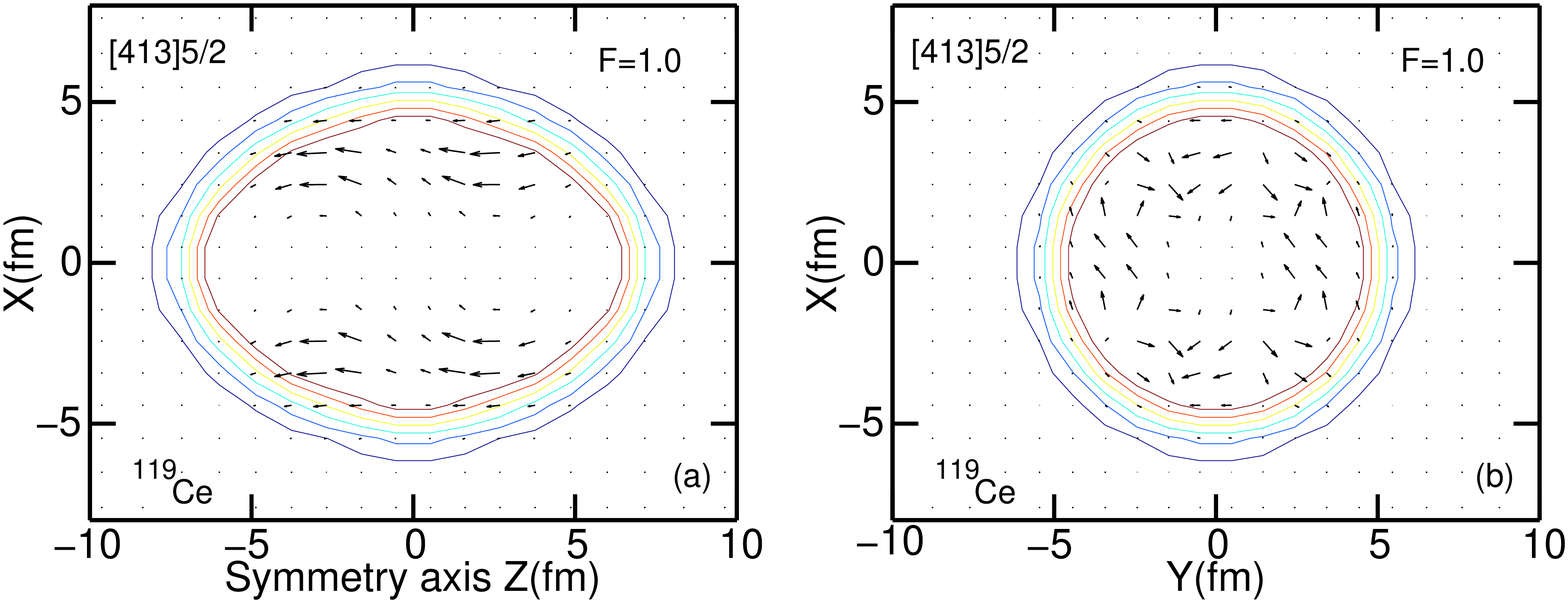}
\vspace{0.0cm}
\vspace{0.0cm}
\caption{(Color online) The same as in Fig.\ \ref{Curr-distr} but for neutron 
current distributions  {\bff j}$^n$({\bff r}) in the $z-x$ plane (at $y=0.48$ fm)
and in the $y-x$ plane (at $z=0.53$ fm) for the $\nu [413]5/2^+$ configuration of
$^{119}$Ce.}
\label{Curr-distr-2}
\end{figure*}

\begin{figure*}
\vspace{0.0cm}
\centering
\includegraphics[width=16.0cm]{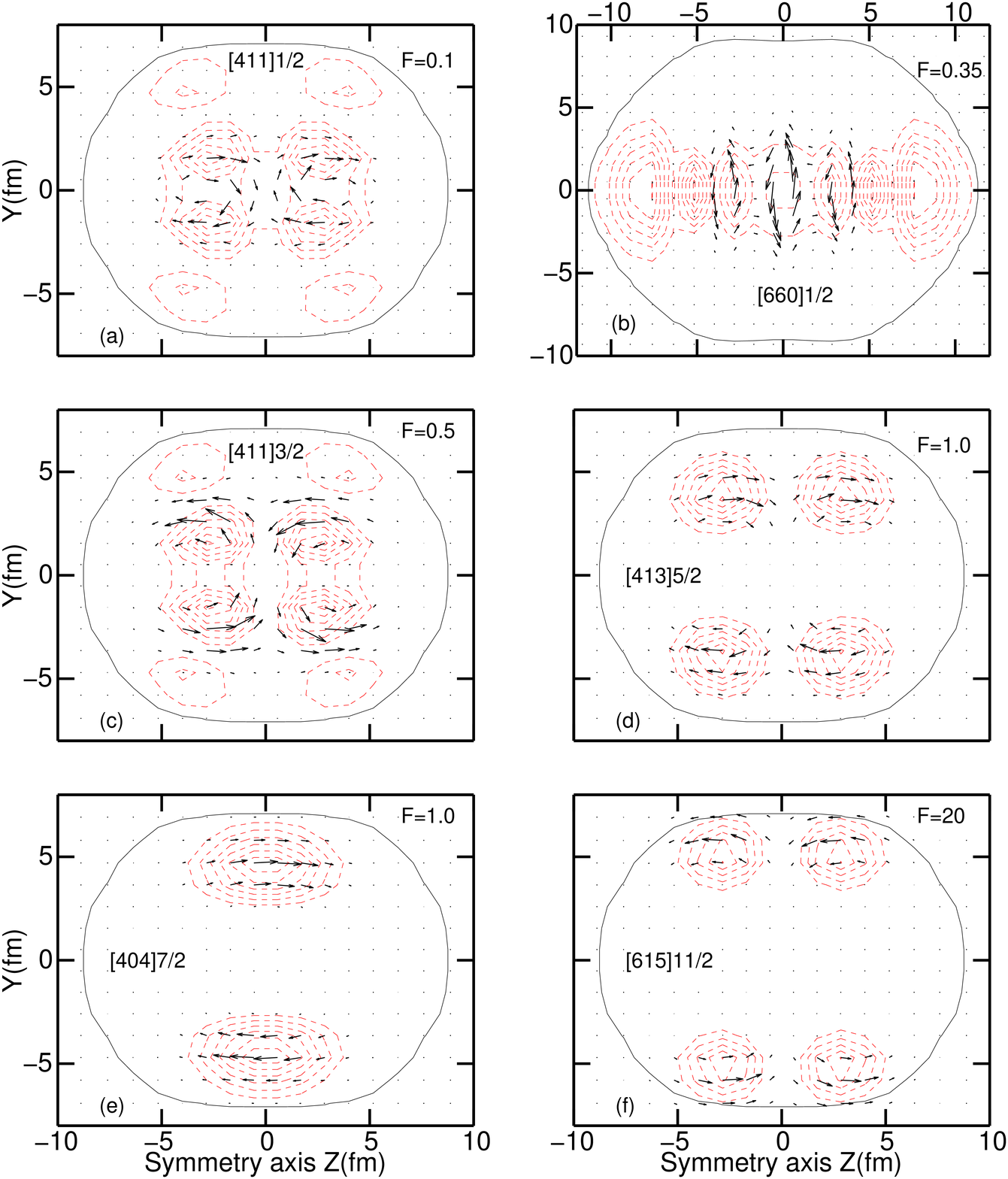}
\vspace{0.0cm}
\vspace{0.0cm}
\caption{(Color online) Current distributions  
{\bff j}$^n$({\bff r}) produced by single neutron in indicated single-particle states
with signature $(r=+i)$ of the $\nu [660]1/2$ configuration in $^{171}$Ce.
The shape and size of the nucleus are indicated by density line which is plotted at 
$\rho=0.01$ fm$^{-3}$ ($\rho=0.0005$ fm$^{-3}$ in panel (b)). The single-neutron density 
distribution due to the occupation of the indicated Nilsson state is plotted starting with 
$\rho=0.0005$ fm$^{-3}$  in step of  0.0005 fm$^{-3}$ ($0.0003$ fm$^{-3}$ in panels 
(d-f)). The currents in panel (d) are plotted at arbitrary units for better visualization. 
The currents in other panels are normalized to the currents in panel (d) by using factor F.  
The currents and densities are shown in intrinsic frame in the $y-z$ plane at $x=0.48$ fm.
}
\label{Curr-distr-single}
\end{figure*}

\subsection{Current distributions}

When discussing current distributions, it is important to remember that the
calculations are performed in one-dimensional cranking approximation. Although the 
rotational frequency is equal to zero in the calculations, the results for the currents 
still obey the symmetries imposed by the cranking approximation. This is clearly seen 
when considering the signature quantum number in the limit of vanishing rotational 
frequency $\Omega_x$ (see Ref.\ \cite{S.book}). In this case definite relations exist 
between the states  $|\nu, r_{\nu}>$ of good signature $r_{\nu}$ ($\nu$ denotes the 
set of additional quantum numbers) and the single-particle states  employed 
usually in the low spin limit. For the latter states in axially symmetric nuclei, 
we obtain doubly degenerate single-particle states $|\nu, \Omega_\nu>$ and 
$|\nu, \bar{\Omega}_\nu>$, where $\Omega_\nu$ denotes the projection of angular 
momentum on the symmetry axis. Here, $|\nu, \Omega_\nu>$ is an eigenstate with 
definite angular momentum projection $\Omega_\nu$, while $|\nu, \bar{\Omega}_\nu>$ 
denotes the time-reversed state (with angular momentum projection $-\Omega_\nu$). In 
the limit of  vanishing rotational frequency $\Omega_x=0$, the states $|\nu, r_{\nu}>$ with 
definite signature $r_{\nu}$ become linear combinations of the states $|\nu, \Omega_\nu>$ 
and $|\nu, \bar{\Omega}_\nu>$
\begin{eqnarray}
|\nu, r_\nu=-i> & =& \frac{1}{\sqrt{2}}\{-|\nu, \Omega_\nu> + (-1)^{\Omega_{\nu}-1/2} |\nu, \bar{\Omega}_\nu>\}, 
\nonumber \\
|\nu, r_\nu=+i> & = & \frac{1}{\sqrt{2}}\{(-1)^{\Omega_{\nu}-1/2}|\nu, \Omega_\nu> +|\nu, \bar{\Omega}_\nu>\} 
\nonumber \\
\end{eqnarray}
 These relations may be considered as a transformation between two representations of the 
single-particle states: the one with good projection $\Omega_\nu$ (the $|\nu, \Omega_\nu>$
representation) and the other with good signature $r$ (cranking representation). In the 
$|\nu, \Omega_\nu>$-representation the alignment of angular momentum vector of a particle 
is specified along the axis of symmetry. As a result, the axial symmetry is conserved and 
only azimuthal currents with respect of the symmetry axis are present. In the cranking 
formalism (which allows also triaxial shapes), the alignment of angular momentum vector of 
a particle is specified along the $x-$axis perpendicular to the axis of symmetry. As a result, 
the currents follow the symmetries of cranking approximation and have the distributions  
discussed below. 

   Total neutron current distributions in the configurations of selected nuclei 
having similar quadrupole deformations are shown in Figs.\ \ref{Curr-distr} and 
\ref{Curr-distr-2}.  They are predominantly defined by the currents generated by 
the blocked orbitals.  This is clearly visible from the comparison of Figs.\ \ref{Curr-distr} 
and \ref{Curr-distr-single}: the latter figure shows the currents produced by single 
neutron in different Nilsson states of the $\nu [660]1/2$ configuration in $^{171}$Ce.
Neutron currents are characterized by the complicated patterns in different 
cross-sections of the nucleus. Fig.\ \ref{Curr-distr-single} clearly shows that
these patterns are defined by the density distributions of the blocked states.
Moreover, there are clear correlations between the patterns of the currents in the $y-z$ 
plane ($z$ is the symmetry axis and $x$ is the rotation axis in the CRMF theory) and the 
$\Omega$ value of the Nillson label of the blocked orbital (Figs.\ \ref{Curr-distr} and 
\ref{Curr-distr-single}). At $\Omega=1/2$, the single-particle  densities are concentrated 
in the vicinity of the axis of symmetry, and, as a consequence, the currents show circulations  
(vortices) which are concentrated in the central region of the 
nucleus. However, with increasing $\Omega$, the densities (and, as a consequence,
the currents) are pushed away from the axis of symmetry of the nucleus towards the 
surface area. In addition, the strength of the currents correlates with $\Omega$. 
As follows from the values of factor $F$ the strongest currents appear for the $\Omega=1/2$ 
states. These orbitals are aligned with the axis of rotation ($x$-axis) already at no rotation. 
As a result, the single-particle angular momentum vector of the $\Omega=1/2$ orbitals performs 
the precession around the $x$-axis, thus orienting the currents predominantly in the $y-z$ plane. 
This extra mechanism is not active in other configurations. The strength of the currents 
decreases with the increase of $\Omega$. For example, the currents generated by the blocked 
$\Omega=11/2$ orbitals are weaker by a factor of almost 200 than the currents generated by the 
blocked $\Omega=1/2$ orbitals (compare scaling factors F  in Fig.\ \ref{Curr-distr} for the 
blocked states with different $\Omega$ values).

  In the $x-y$ plane, the majority of the configurations show the current pattern 
(although with different strength of the currents and their localization in space) 
visible on Fig. \ref{Curr-distr-2}b, while the typical pattern of the currents in the 
$x-z$ plane is shown in Fig.\ \ref{Curr-distr-2}a. Figs.\ \ref{Curr-distr}d,e show 
that the change of the signature of the blocked orbital leads only to a change in the 
direction of the currents.

\subsection{Particle number dependences of additional binding due to NM}
\label{Mass-dep-en}

\begin{figure*}
\centering
\includegraphics[width=16.0cm]{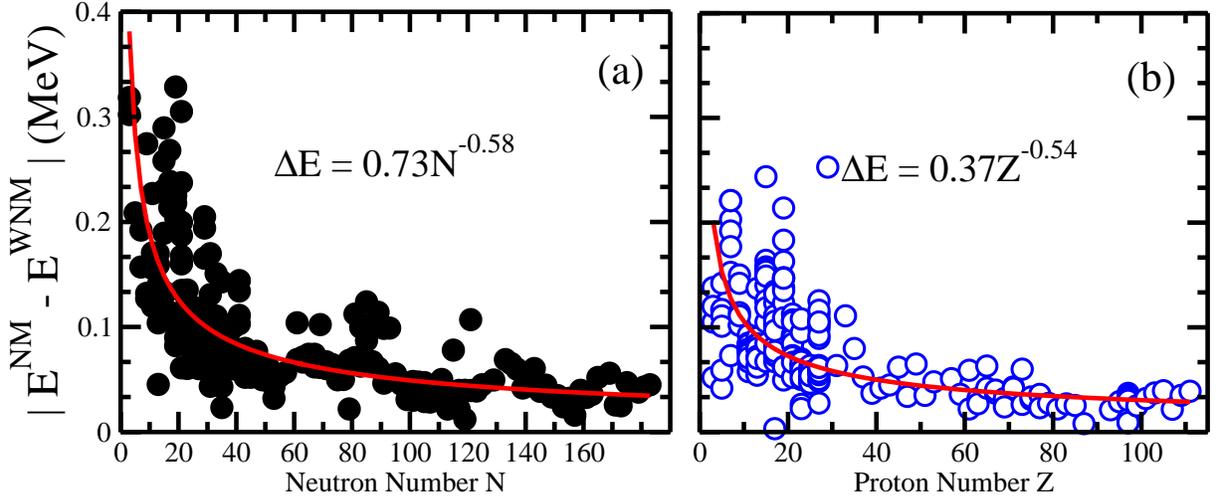}
\vspace{0.8cm}
\caption{(Color online) 
Neutron and proton dependences of additional binding due to NM. Open  and solid 
circles are used for odd-proton and odd-neutron nuclei, respectively.}
\label{Part-dep}
\end{figure*}

  Neutron and proton number dependences of additional binding due to NM 
(the $|E^{NM}-E^{WNM}|$ quantity) are presented in Fig.\  
\ref{Part-dep}. These figures are based on the results obtained in Sects.\ 
\ref{Bind-light} and \ref{Ce-isotopes} and on some extra calculations.
These extra calculations include odd-$Z$ nuclei with $N=94$, 
odd-$N$ nuclei with $Z=98$, and odd-$Z$ nuclei with $N=154$ and cover
these isotope and isotone chains from proton- to neutron-drip lines. 

   The calculations of nuclei around $^{249}$Cf were also performed in order 
to check the impact of pairing on the $(E^{NM}-E^{WNM})$ quantity. For the 
same blocked states, the $(E^{NM}-E^{WNM})$ values obtained in the calculations 
without pairing in the present manuscript were compared with the ones obtained in 
the Relativistic Hartree-Bogoliubov calculations of Ref.\ \cite{A250} (see Table 
IV of Ref.\ \cite{A250}).  Although the pairing decreases additional binding due 
to NM in the most of the cases, there are still one-(quasi)particle configurations 
in which the $|E^{NM}-E^{WNM}|$ quantity is smaller in the calculations without 
pairing. This is a consequence of the complicated nature of the $E^{NM}-E^{WNM}$ 
quantity defined by (i) the interplay of time-odd mean fields and the polarization 
effects (Sect.\ \ref{Mech-bind-to}) and by (ii) the differences  in the impact of 
pairing on different terms of total energy.

  The calculated $|E^{NM}-E^{WNM}|$ quantities were fitted by simple parametrization
\begin{eqnarray}
\Delta E = \frac{c}{Q^{\alpha}}
\label{Delta}
\end{eqnarray}
where $Q$ is equal either to proton $Z$ or neutron $N$ numbers. Note
that the $|E^{NM}-E^{WNM}|$ values from odd-proton (odd-neutron) nuclei were
used in the fit of $Z-(N-)$dependence of $\Delta E$. The results of the fits are 
shown by solid lines in Fig.\ \ref{Part-dep}. One can see that the powers 
$\alpha$  are similar for different fits (proton or neutron). On the other 
hand, the magnitudes $c$ differ considerably between proton and neutron 
quantities, indicating weaker additional binding due to NM for odd-proton
nuclei. This result is consistent with the analysis obtained in Sect.\ \ref{Sect-odd-odd}.

The three-point indicator \cite{DMNSS.01}
\begin{eqnarray}
\Delta ^{(3)}(N) = \frac{\pi_N}{2} \left[ B(N-1) + B(N+1) - 2 B(N) \right]
\label{neut-OES}
\end{eqnarray} 
is frequently used to quantify the odd-even staggering (OES) of binding energies. 
Here $\pi_N=(-1)^N$ is the number parity and $B(N)$ is the (negative) binding energy 
of a system with $N$ particles. In Eq.\ (\ref{neut-OES}), the number of protons $Z$
is fixed, and $N$ denotes the number of neutrons, i.e. this indicator gives the 
neutron OES. The factor depending on the number parity $\pi_N$ is chosen so that 
the OES centered on even and odd neutron number $N$ will both be positive. An analogous 
proton OES indicator $\Delta ^{(3)}(Z)$ is obtained by fixing the neutron number $N$ and 
replacing $N$ by $Z$ in Eq.\ (\ref{neut-OES}).

 The $\Delta ^{(3)}(N)$ (and similarly $\Delta ^{(3)}(Z)$) quantity will be modified 
in the presence of time-odd mean fields as
\begin{eqnarray}
\Delta ^{(3)}_{TO}(N) = \Delta ^{(3)}_{WTO}(N) + \delta E_{TO}  
\end{eqnarray}
where  the subscripts 'TO' and 'WTO' indicate the values obtained in the calculations 
with and without time-odd mean fields, and $\delta E_{TO}$ is the contribution coming 
from time-odd mean fields. If the $\Delta ^{(3)}(N)$ quantity is centered at odd-$N$ 
nucleus, the $\delta E_{TO}$ quantity represents the change of binging energy of this
odd-mass nucleus induced  by time-odd mean fields.  This is because the time-odd mean 
fields have no effect on the binding energies of the ground states of even-even nuclei. Note 
that with such selection $\delta E_{TO}$ is negative if the time-odd mean fields provide 
additional binding in odd-mass nucleus.

  In the CDF theory, the $\delta E_{TO}$ quantity is equal to $E^{NM}-E^{WNM}$ and
thus it is always negative: this result does not depend on the RMF parametrization
(see Sect.\ \ref{Bind-light} for the dependence of the $E^{NM}-E^{WNM}$ quantity on 
the RMF parametrization). In addition, the magnitude of the $\delta E_{TO}$ quantity 
depends only weakly on the RMF parametrization. On the contrary, the sign and the 
magnitude of $\delta E_{TO}$ depends strongly on the parametrization in the Skyrme EDF 
calculations. For example, in the calculations with the SLy4 force the 
$\delta E_{TO}$ quantity is positive  for medium mass nuclei (Refs.\  \cite{BBNSS.09,DBHM.01,DBHM.01-1}) 
but negative in light nuclei (Ref.\ \cite{S.99}). On the other hand, the $\delta E_{TO}$
quantity will be positive in light nuclei in the calculations with the SIII 
parametrizations \cite{S.99}. 

   It is interesting to compare the averaged effects of time-odd mean fields as 
given by the $\Delta E$ quantity with the experimental global trends for OES as 
shown by dashed lines in Fig.\ 2 in Ref.\ \cite{BBNSS.09}. The latter trends were 
obtained using phenomenological parametrization with the same functional dependence as 
in Eq.\ (\ref{Delta}) with $c=4.66$ MeV (4.31 MeV) and $\alpha=0.31$ for neutron (proton)
data sets. The comparison of theory and experiment suggests that time-odd mean field
contributions into OES can be as large as 10\% in light systems and around 5-6\% in
heavy systems. These are non-negligible contributions which have to be taken into
account when the strength of pairing interaction is defined from the fits to experimental
OES. The analysis of the Sn isotopes in Ref.\ \cite{RBRM.99} showed that time-even 
and time-odd polarization effects induced by odd nucleon produce OES reduced by
about 30\% as compared to the ones obtained in standard spherical calculations. As a 
consequence, an enhancement of pairing strength by about 20\% is required to compensate 
for that effect. Our calculations show much smaller reduction of OES in part because the 
polarization effects in time-even channel are already taken into account in the calculations 
without NM.  Thus, the current calculations suggest that a much smaller increase of the 
strength of pairing (by approximately 5\%) would be required  to compensate for the 
reduction of OES due to time-odd mean fields.

\section{The mechanism of additional binding due to NM in odd-mass nuclei}
\label{Mech-bind-to}

  In the current section, a detailed analysis of the impact of NM on the energies 
of the single-particle states and on different terms in the total energy expression 
(Eq.\ (\ref{Etot})) is performed in order to better understand the microscopic 
mechanism of additional binding due to NM. We use the $\nu [413]5/2$ configuration 
of $^{119}$Ce as an example in this analysis.

\subsection{Energy splittings of time-reversal counterpart single-particle 
states in the presence of NM}
\label{E-split}

\begin{figure}
\centering
\includegraphics[width=7.0cm]{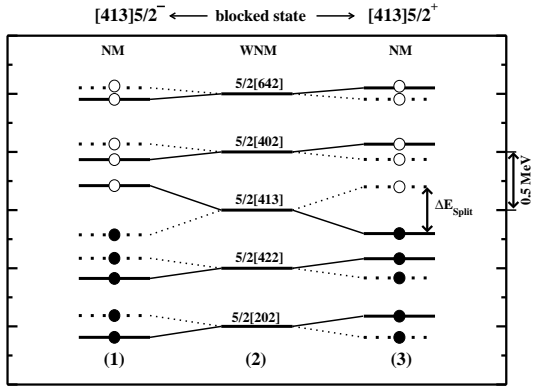}
\vspace{0.8cm}
\caption{(Columns (1) and (3)) The energy splittings $\Delta E_{split}$ between 
different signatures of the single-particle states in the presence of NM. The 
results of the calculations are shown for the configurations of $^{119}$Ce in 
which either the $\nu [413]5/2^-$ (column (1)) or $\nu [413]5/2^+$ (column (3)) 
states are blocked. These signatures are degenerated in energy in the calculations 
without NM (column (2)). Note that the single-particle states of interest are shown 
at arbitrary absolute energy in the column (2). Solid (open) circles indicate 
occupied (unoccupied) states. Solid (dotted) lines are used for the $r=+i$ 
($r=-i$) states. }
\label{Esp-splitting}
\end{figure}

 Fig.\ \ref{Esp-splitting} shows that the presence of time-odd mean fields 
leads to the energy splitting $\Delta E_{split}(i)$ of the single-particle states which 
are time-reversal counterparts. This corresponds to the removal of the Kramer's 
degeneracy of these states. One of these states moves up by $\approx \Delta E_{split}/2$ 
as compared with its position in the absence of NM, while another moves down by 
$\approx \Delta E_{split}/2$.  

 Detailed analysis of the single-particle spectra in $^{119}$Ce and $^{123}$Xe reveals 
general features which are also found in other nuclei. The $^{119}$Ce nucleus is axially 
symmetric ($\gamma=0^{\circ}$) while $^{123}$Xe is triaxial with $\gamma=-26^{\circ}$. 
This difference in the symmetry of nucleus results in important consequences: the 
energy splittings appear in all single-particle states in triaxial nuclei, while 
only the states with $\Omega=\Omega_{bl}$ (the subscript {\it 'bl'} indicates the blocked 
state)
experience such splittings in  axially 
symmetric nuclei. The former feature is due to the fact that $\Omega$
is not a good quantum number in triaxial nuclei and each single-particle state
represents a mixture of the basic states with different values of  $\Omega$.

  It is important to mention that the occupied and unoccupied states as well 
as the proton and neutron states show energy splittings (Fig.\ \ref{Esp-splitting}).
The splittings of the proton and neutron states of the same structure are similar. 
This is because the largest contribution to magnetic potential (Eq.\ (\ref{magnetic})) 
is due to space-like components of the $\omega$-meson fields which do not depend 
on the isospin. In addition, the occupied state is always more bound than its 
unoccupied time-reversal counterpart.

  The change of the signature of the blocked state leads to the inversion of the 
signatures in all pairs of time-reversal orbitals (compare columns (1) and (3) 
in Fig.\ \ref{Esp-splitting}). The explanation for this process is the 
following.  The change of the signature of the blocked state results in the 
change of the direction of the currents to the opposite one (compare Fig.\ 
\ref{Curr-distr}d with Fig.\ \ref{Curr-distr}e). This leads to the change of 
the direction of the vector  potential ${\bff V}({\bff r})$ in the Dirac equation 
to the opposite one, which in turn causes the inversion of the signatures in all 
pairs of time-reversal orbitals.  However, the additional  binding due to NM 
(the $E^{NM}-E^{WNM}$ quantity) does not depend on the signature of the blocked 
state in odd-mass nuclei.

\subsection{Polarization effects induced by NM}
\label{Pol-odd}

 The polarization effects induced by NM are investigated by considering 
its impact on different terms of the total energy (Eq.\ (\protect\ref{Etot})). 
The results of this study are shown in Table \ref{NM-odd}. One can see that
the total energy terms can be split into two groups dependent on how they are 
affected by NM. The first group includes  the $E_{\sigma NL}$, $E_{\rho}^{TL}$, 
$E_{\rho}^{SL}$ and $E_{Coul}$ terms which are only weekly influenced by NM, and 
thus, they will not be discussed in detail. 

\begin{table}[h]
\caption{The impact of NM on different terms of the total energy (Eq.\ (\protect\ref{Etot})) 
in the $[413]5/2^+$ configuration of $^{119}$Ce. Second column shows the absolute energies 
[in MeV] of different energy terms in the case when NM is neglected. Third and forth columns 
show the changes $\Delta E_{i}=E^{NM}_{i}-E^{WNM}_{i}$ [in MeV] in the energies of these 
terms induced by NM in self-consistent (column 3) and perturbative (column 4) calculations.
Note that we show only nonzero quantities in column 4. The relevant quantities are also 
shown for kinetic energy $E_{kin}$ (Eq.\ (\ref{Ekin-f})) in the last line.} 
\label{NM-odd}
\vspace{0.5cm}
\begin{center}
\begin{tabular}{|c|c|c|c|c|} \hline
Quantity            &  $E_i^{WNM}$  & $\Delta E_i$ & $\Delta E_i^{pert}$ \\ \hline
     1              &      2       &       3            &      4    \\ \hline
$E_{part}$          & $-2849.889$   &  $-0.410$           &  $-0.237$ \\
$E_{\sigma}$        &  $-17079.532$ &  $-2.231$         &     \\
$E_{\sigma NL}$     &   343.341      &  $-0.017$             &  \\
$E_{\omega}^{TL}$      &  14356.156  &  $2.054$           &  \\
$E_{\omega}^{SL}$      &  0.0        &  $-0.124$            &  $-0.124$    \\
$E_{\rho}^{TL}$        &  2.044      &  $0.003$            &    \\
$E_{\rho}^{SL}$        &  $ 0.0$     &  $-0.010$        &  $-0.010$       \\
$E_{Coul}$          &  $ 481.196$   &  $0.017$          &              \\
$E_{cm}$            &  $ -6.252 $   &  0.0              &             \\
$E_{tot}$           &  $ -959.349$  & $-0.104$         & $-0.103$    \\ \hline \hline
$E_{kin}$           &  $1630.386$   & $-0.099$          & $-0.237$     \\ \hline
\end{tabular}
\end{center}
\end{table}

  The second group is represented by the $E_{part}$, $E_{\sigma}$, $E_{\omega}^{TL}$ 
and $E_{\omega}^{SL}$ terms which are strongly affected by NM.  The $E_{\omega}^{SL}$ 
term is directly connected with the nucleonic currents (see Eq.\ (\ref{E-s-omega})). 
The $E_{\sigma}$ and $E_{\omega}^{TL}$ terms depend only indirectly on time-odd mean 
fields: the minimization of the total energy in the presence of time-odd terms leads 
to a very small change of equilibrium deformation induced by NM. The quadrupole and 
hexadecapole moments change by $10^{-4}$ of their absolute value when the NM is 
switched on; a similar magnitude of changes is seen also in $E_{\sigma}$ and $E_{\omega}^{TL}$. 
One should keep in mind that only the $E_{\sigma}+E_{\omega}^{TL}$ quantity has a deep 
physical meaning since it defines a nucleonic potential; this sum is modified by NM only on 
-177 keV.

\begin{table}[h]
\caption{The same as in Table \ref{NM-odd} but for the $[606]13/2^+$ configuration in 
$^{183}$Ce.}
\label{NM-odd-3}
\vspace{0.5cm}
\begin{center}
\begin{tabular}{|c|c|c|c|} \hline
Quantity            &  $E_i^{WNM}$    & $\Delta E_i$      &  $\Delta E_{i}^{pert}$      \\ \hline
     1              &      2         &       3           &         4                  \\ \hline
$E_{part}$           & $-4139.512$    &  $-0.157$         &   $-0.095$                 \\
$E_{\sigma}$         &  $-23720.872$  &  $-0.608$        &                             \\
$E_{\sigma NL}$      &   541.423      &  $-0.004$          &                            \\
$E_{\omega}^{TL}$    &   19696.151    &  $0.538$           &                            \\
$E_{\omega}^{SL}$    &     0.0        &  $-0.043$         &     $-0.043$                 \\
$E_{\rho}^{TL}$      &  236.863       &  $0.009$        &                               \\
$E_{\rho}^{SL}$      &    $ 0.0$      &  $-0.005$      &   $-0.005$                    \\
$E_{Coul}$          &  $437.326 $    &  $0.002$          &                               \\
$E_{cm}$            &  $ -5.416 $    &  0.0             &                               \\
$E_{tot}$           &  $-1335.818 $  & $-0.045$         &       $-0.047$                \\ \hline \hline
$E_{kin}$           &  $2561.558$    & $-0.045$         &     $-0.095$                  \\    \hline
\end{tabular}
\end{center}
\end{table}

  The largest modification (by $-410$ keV) is seen in the $E_{part}$ energy, with 
the half of it coming from the change of the single-particle energy 
(by $\approx -200$ keV) of the blocked orbital (the $\nu [413]5/2$ orbital)
in the presence of NM.  Note that since both signatures of other pairs of 
time-reversal orbitals below the Fermi level are occupied, the large energy 
splittings $E_{split}$ seen for some of them do not have a considerable impact on 
$E_{part}$ (see Eq.\ (\ref{Epart})) since this splitting is nearly symmetric with 
respect to the position of these orbitals in the absence of NM. Thus, the rest of 
the modification of $E_{part}$ is related to small changes in the single-particle 
energies of occupied states caused by the changes in the equilibrium deformation 
induced by NM.

  This detailed analysis clearly indicates that the $E^{NM}-E^{WNM}$ quantity is 
defined by both time-odd mean fields and the polarization effects in time-even 
mean fields induced by time-odd mean fields. $E^{NM}-E^{WNM}=-104$ keV is a result 
of near cancellation of the contributions due to fermionic ($-410$ keV) and mesonic 
($-306$ keV) degrees of 
freedom.  Note that the latter appears with a negative sign in Eq.\ (\ref{Etot}).
The fermionic degrees of freedom are represented by the $E_{part}$ and $E_{cm}$
terms, while the other terms of the total energy are related to the mesonic degrees 
of freedom. The fermionic contribution into  $E^{NM}-E^{WNM}$ is defined by more 
or less equal contributions from time-odd mean fields and the polarization effects 
in time-even fields. On the contrary, time-odd mean fields define only 
$\approx 1/3$ ($E_{\omega}^{SL}=-0.124$ keV) of the mesonic contribution into 
$E^{NM}-E^{WNM}$, while the rest is due to polarization effects in time-even 
mean fields.

\begin{figure}
\centering
\vspace{0.5cm}
\includegraphics[width=7.7cm]{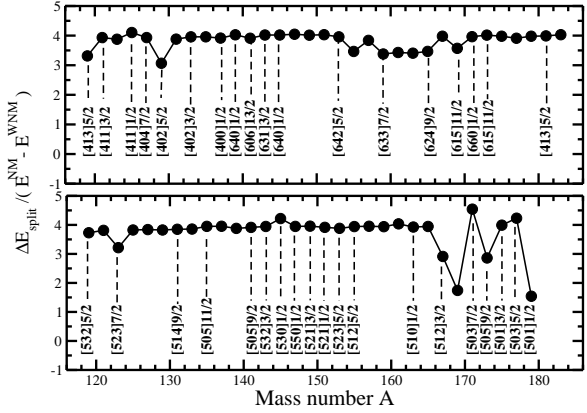}
\vspace{0.5cm}
\caption{ The ratio $\Delta E_{split}/(E^{NM}-E^{WNM})$ in the Ce isotopes. 
The structure of the blocked states is shown by the Nilsson labels; the states 
at and to the right of the Nilsson label up to the next Nilsson label have 
the same blocked state. }
\label{Ce-ratio}
\end{figure}

  It turns out that these contributions are highly correlated as can be seen 
from the ratio $\Delta E_{split}/(E^{NM}-E^{WNM})$ in the Ce isotope chain (Fig.\ 
\ref{Ce-ratio}). $\Delta E_{split}$ depends only on time-odd mean fields in 
fermionic channel, while $(E^{NM}-E^{WNM})$ depends both on time-odd 
mean fields and the polarizations effects  in time-even mean fields in fermionic 
and mesonic channels. One can see that $\Delta E_{split}/(E^{NM}-E^{WNM})\approx 4$ 
for the majority of nuclei. Similar relation exists also in the Skyrme EDF calculations 
for the  Ce isotopes (see Eq.\ (7) in Ref.\ \cite{DBHM.01}).

\begin{figure*}
\centering
\vspace{0.5cm}
\includegraphics[width=16.0cm]{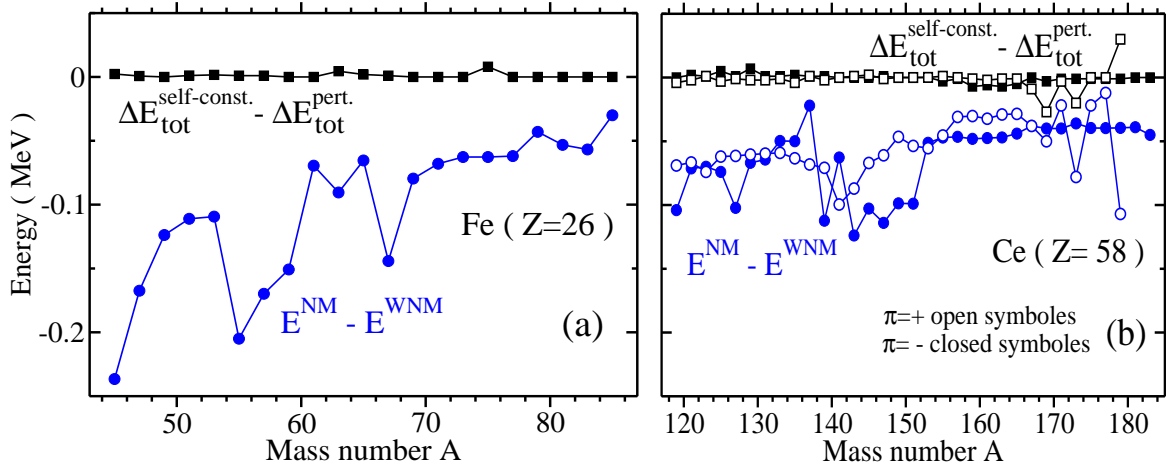}
\caption{The $\Delta E^{self-const}_{tot} - \Delta E^{pert}_{tot}$ and $E^{NM}-E^{WNM}$ 
quantities for odd-neutron Fe and Ce nuclei.}
\label{DeltaE-E}
\end{figure*}

  The impact of NM on different terms of the total energy in the $\nu [606]13/2^+$ 
configuration of the $^{183}$Ce nucleus, which is located at the neutron drip line, is 
shown in Table \ref{NM-odd-3}. The comparison of Tables \ref{NM-odd} and \ref{NM-odd-3} 
allows to understand the microscopic origin of general trend which shows the decrease 
of the impact of NM (reflected in the $(E^{NM}-E^{WNM})$ quantity) with increasing particle 
(proton, neutron or mass) number (see Figs.\ \ref{light-A} and \ref{Part-dep}). In the 
$^{183}$Ce nucleus, the impact of NM on the $E_{\omega}^{SL}$ and $E_{\rho}^{SL}$ 
terms, which directly depend on time-odd mean fields, decreases by the factors close 
to 3 and 2 relative to the $^{119}$Ce case (see Table \ref{NM-odd}), respectively.
The impact of NM on the $E_{\sigma}$, $E_{\sigma NL}$, $E_{\omega}^{TL}$, $E_{\rho}^{TL}$, 
$E_{Coul}$ terms, which depend on time-odd mean fields only through polarization effects, 
decreases even more dramatically  (by a factor close to 4). Note that the contribution 
of the $\rho$-meson to the $(E^{NM}-E^{WNM})$ quantity is marginal even at the neutron-drip 
line. Other investigated cases also indicate the decrease of the impact of NM with 
increasing particle number. 

  The general trend of the decrease of the impact of NM on binding energies
with increasing particle number can be understood in the following way.  The effects 
attributable to NM are produced by odd particle which breaks time-reversal symmetry. 
With increasing particle (proton, neutron of mass) number the nucleus becomes larger 
and thus more robust towards time-odd and polarization effects induced by odd 
particle (or in other words, the effective impact of single particle on total nuclear 
properties becomes smaller).

 It is interesting to compare the results of self-consistent and perturbative calculations. The 
$\Delta E_{i}=E^{NM}_{i}-E^{WNM}_{i}$ quantities will be used for simplicity in further discussion.
These quantities are shown in columns 3 and 4 of Tables \ref{NM-odd} and \ref{NM-odd-3}. The 
$\Delta E_{\sigma}$, $\Delta E_{\sigma NL}$,  $\Delta E_{\omega}^{TL}$ and $\Delta E_{\rho}^{TL}$ 
quantities are zero in perturbative calculations. $\Delta E_{Coul}=0$ for odd-neutron system in 
perturbative calculations (Tables \ref{NM-odd} and \ref{NM-odd-3}), but it can differ from 0 in the 
systems containing odd number of protons (Tables \ref{NM-prot} and \ref{Table-Cl34}). The results 
of self-consistent and perburbative calculations for the $\Delta E_{\omega}^{SL}$ and $\Delta E_{\rho}^{SL}$ 
quantities are the same with the exception of conf. B in $^{34}$Cl where only small difference exists 
(Tables \ref{NM-odd}, \ref{NM-odd-3} 
and \ref{Table-Cl34}).

  It is seen from Tables \ref{NM-odd} and  \ref{NM-odd-3} that
\begin{eqnarray} 
\Delta E_{tot}^{self-const} \approx \Delta E_{tot}^{pert}
\label{Eqq}
\end{eqnarray}
for odd-neutron nuclei. Note that the superscript {\it 'self-const'} refers to fully 
self-consistent results. Fig.\ \ref{DeltaE-E} shows that this equality 
is fulfilled in the majority of nuclei of the Fe and Ce isotope chains with high degree 
of accuracy (as compared with the $E^{NM}-E^{WNM}$ quantities). {\it These results clearly 
indicate that the additional binding due NM (the $E^{NM}-E^{WNM}$ quantity) is defined mainly
by time-odd fields and that the polarization effects in fermionic and mesonic sectors of 
the model cancel each other to a large degree.}

 As a consequence it is important to understand the relations between 
different polarization effects. Particle energy $E_{part}^{self-const}$ obtained in self-consistent 
calculations can be split into two parts: the part $E_{part}^{TO}$ which directly depends 
on time-odd mean fields and the part $E_{part}^{pol}$ which is defined by the polarization 
effects in the fermionic sector of model. Thus, $E_{part}^{self-const}=E_{part}^{TO}+E_{part}^{pol}$ and 
$E_{part}^{pert}\approx E_{part}^{TO}$. Taking into account Eq.\ (\ref{Etot}) and above mentioned 
features of the $\Delta E_{i}^{pert}$ terms one can conclude that 
\begin{eqnarray}
\Delta E_{part}^{pol} =   \Delta E_{\sigma}^{self-const} + \Delta E_{\sigma NL}^{self-const} \nonumber \\  
                       + \Delta E_{\omega}^{TL [self-const]} +  \Delta E_{\rho}^{TL [self-const]} \nonumber \\ 
                       + \Delta E_{Coul}^{self-const}.  
\label{Pol-ferm}
\end{eqnarray}
This relation clearly indicates that the polarizations effects in the fermionic ($E_{part}^{pol}$ 
term) and mesonic ($\Delta E_{\sigma}^{self-const}$, $\Delta E_{\sigma NL}^{self-const}$, 
$\Delta E_{\omega}^{TL [self-const]}$, and $\Delta E_{\rho}^{TL [self-const]}$ terms) sectors
of the model are strongly correlated. Eq.\ (\ref{Pol-ferm}) also allows to understand clearly 
the physical origin of $\Delta E_{part}^{pol}$. The terms on right hand side are related to the 
change of the nucleonic potential induced by NM. This change leads to the modifications of the 
single-particle energies of all occupied states (as compared with the case when NM is absent)
which are reflected in $\Delta E_{part}^{pol}$. On the contrary, the $\Delta E_{part}^{TO}$ is due 
to the breaking of the Kramers degeneracy between the blocked state and its unoccupied 
time-reversal counterpart. Note that $\Delta E_{part}^{TO}\approx - 1/2 \Delta E_{split}$ 
(minus sign reflects the fact that the blocked state is always more bound in the presence 
of NM) and the $\Delta E_{split}$ values obtained in self-consistent and perturbative calculations are
the same for the pairs of time-reversal counterpart states involving blocked state.

\begin{figure}
\centering
\vspace{0.5cm}
\includegraphics[width=8.0cm]{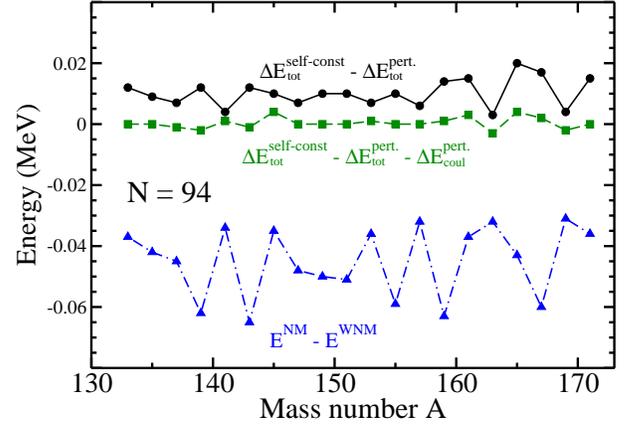}
\caption{The same as in Fig.\ \ref{DeltaE-E}, but for odd-proton $N=94$ nuclei.}
\label{DeltaE-Eprot}
\end{figure}

 The relation similar to Eq.\ (\ref{Eqq}) exists also in odd-proton nuclei but
in this case it has to be corrected for the $\Delta E_{Coul}^{pert}$ energy change: 
\begin{eqnarray}
\Delta E_{tot}^{self-const} \approx \Delta E_{tot}^{pert} + E_{Coul}^{pert}.
\label{OOOT}
\end{eqnarray}
Fig.\ \ref{DeltaE-Eprot} shows that this relation is fulfilled in odd-proton $N=94$ 
nuclei with high degree of accuracy (as compared with the  $E^{NM}-E^{WNM}$ quantities). 
Eq.\ (\ref{OOOT}) also leads to the condition of Eq.\ (\ref{Pol-ferm}) and to the 
interpretation of $E_{part}^{pol}$ discussed above.

\section{The impact of time-odd mean fields on the properties of
proton-unstable nuclei}
\label{Sect-proton}

\begin{figure}
\centering
\includegraphics[width=8.0cm]{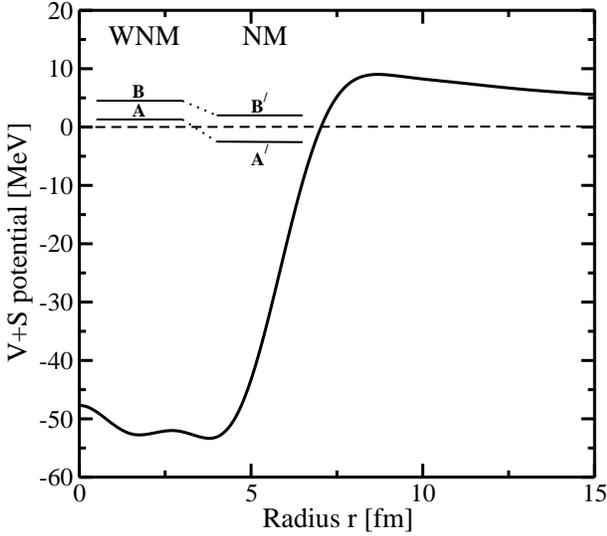}
\vspace{0.8cm}
\caption{Schematic illustration of the impact of time-odd mean fields on 
the properties of odd-proton nuclei in the vicinity of the proton drip line. 
The single-proton states, involved in the mechanisms discussed in the text, 
and proton  nucleonic potential (which also includes the Coulomb potential) 
are shown in the figure.}   
\label{Mech-proton}
\end{figure}

The blocked state always has lower energy than its unoccupied time-reversal 
counterpart in the calculations with NM; this fact does not depend on the signature 
of the blocked state (Sect.\ \ref{E-split}). The energy of the blocked state in the 
presence of NM is lower by $\approx \Delta E_{split}/2$ than the energy of the
same state in the absence of NM. This additional binding will affect the 
properties of the nuclei in the vicinity of the proton-drip line via two mechanisms 
discussed below. They are schematically illustrated in Fig.\ \ref{Mech-proton}.

 In the first mechanism, the nucleus, which is proton  unbound (state A in Fig.\ 
\ref{Mech-proton}) in the calculations without NM, becomes proton bound in the 
calculations with NM (state A' in Fig.\ \ref{Mech-proton}). The necessary condition 
for this mechanism to be active is the requirement that the energy of the single-particle 
state in the absence of NM is less than $\Delta E_{split}/2$. This mechanism can be 
active both in the ground and excited states of the nuclei in the vicinity of the 
proton-drip line.

  In the second mechanism, the energy of the single-particle state (state B' 
in Fig.\ \ref{Mech-proton}) is lower in the presence of NM, but the state still 
remains unbound. This will affect the decay properties of proton emitters and 
the possibilities of their observation. Indeed, the lowering of the energy of 
the single-proton state will decrease the probability of emission of the proton 
through combined Coulomb and centrifugal barrier. Many results of the physics
of proton emitters are conventionally expressed in terms of the $Q_p$ energies
which depend on the difference of the binding energies of parent (odd-proton) and 
daughter (even-proton) nuclei. Note that for simplicity we consider here only 
even-$N$ nuclei. NM leads to an additional binding in odd-proton nucleus but 
it does not affect the binding of even-proton nucleus. Thus, the $Q_p$ values 
are lower by the value of this additional binding when the NM is taken into 
account.

  Two consequences follow from lower $Q_p$ values. First, experimental observation 
of proton emission from the nucleus will become impossible if the $Q_p$ value 
moves outside the $Q_p$ window favorable for the observation of proton emission 
or becomes possible if the $Q_p$ value moves into the $Q_p$ window favorable for 
the observation of proton emission. The size 
of the $Q_p$ window for rare-earth proton emitters is about $0.8-1.7$ MeV, while it is 
much smaller in lighter nuclei \cite{WD.97,ASN.97}. Large $Q_p$ values outside this 
window result in extremely short proton-emission half-lives, which are difficult to 
observe experimentally. On the other hand, the decay width is dominated by $\beta^+$ 
decay for low $Q_p$ values below the $Q_p$ window. This consequence of the lowering of
$Q_p$ due to NM is especially important in light nuclei where the impact of NM on 
binding energies is especially pronounced and the $Q_p$ window is narrow.

  Second, the lowering of the $Q_p$ values due to NM will increase the half-lives of 
proton emitters. For example, the lowering of $Q_p$ due to NM will be around 50 keV 
in rare-earth region since this is typical value of additional binding due to NM in 
odd-mass nuclei of this region (Sect.\ \ref{Ce-isotopes}). This can increase the half-lives 
of proton emitters by a factor of $\approx 2$ at the upper end of the $Q_p$ window and 
by a factor of $\approx 4$ at the bottom end of the $Q_p$ window (see Fig.\ 5 in Ref.\ 
\cite{ASN.97}). The effects of NM have been neglected in the existing RHB studies of 
proton emitters with $Z\geq 50$ (see, for example, Ref.\ \cite{LVR.01}) but this 
should not introduce significant error in this mass region.

  On the other hand, the impact of NM can be dramatic on the half-lives of proton 
emitters in lighter nuclei. This is due to two factors, namely, (i) the general increase 
of additional binding due to NM  and the magnitude of $\Delta E_{split}$ with decreasing
mass and (ii) the narrowing of the $Q_p$ window with the decrease of mass due to the 
lowering of the Coulomb barrier. This can be illustrated by several examples.
The change in proton energy of around 300 keV in $^{69}$Br causes a change in the 
proton decay lifetime of 11 orders of magnitude \cite{WD.97}. This effect is even 
more pronounced in lighter systems. The half-life window of 10 to $10^{-4}$ s corresponds 
to proton energies of $100-150$ keV in nuclei around $Z=20$ \cite{G.60}, while the 
variation of the $Q_p$ value between 3 to 50 keV in $^7$B changes the half-lives by 
30 orders of magnitude \cite{ASN.97}. The energy changes quoted in these examples 
are either of similar magnitude or even smaller as compared with the changes of the 
energies of single-proton states and the $Q_p$ values induced by NM. As a result, one 
can conclude that the effects of time-odd mean fields have to be taken into account when 
attempting to describe the properties of proton emitters in light nuclei. 

\section{Odd-odd mass nuclei: a model study of the impact of nuclear
magnetism on binding energies.}
\label{Sect-odd-odd}

\begin{figure}
\centering
\includegraphics[width=8.0cm]{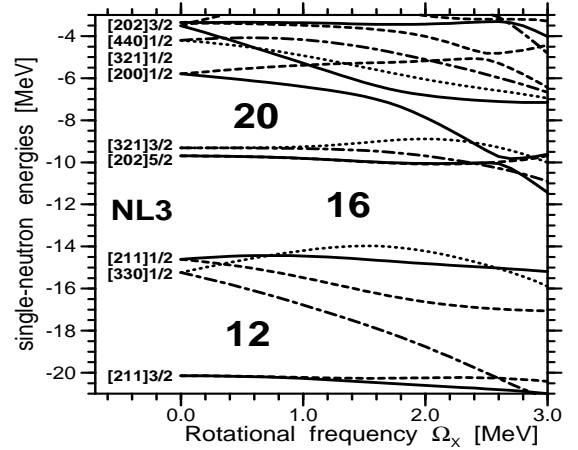}
\caption{Neutron single-particle energies (routhians) as a function of the 
rotational frequency $\Omega_x$. They are given 
along the deformation path of the doubly-magic SD configuration $\pi 3^2 \nu 3^2$ 
in $^{32}$S. Solid, short-dashed, dot-dashed and dotted lines indicate 
$(\pi=+,\,\,r=-i)$, $(\pi=+,\,\,r=+i)$, $(\pi=-,\,\,r=+i)$ and 
$(\pi=-,\,\,r=-i )$ orbitals, respectively.  At $\Omega_x=0.0$ MeV, the 
single-particle orbitals are labeled by means of the asymptotic quantum 
numbers $[Nn_z\Lambda]\Omega$ (Nilsson quantum numbers) of the dominant
component of the wave function.}
\label{routh-s32}
\end{figure}

\begin{figure}
\centering
\includegraphics[width=8.0cm]{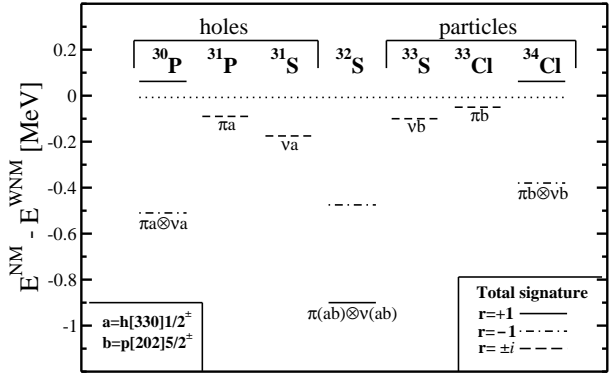}
\caption{The impact of NM on binding energies of different configurations 
under study.  The energies of the configurations calculated without NM are 
normalized to zero. To guide an eye the dotted line is used for zero value of the
$E^{NM}-E^{WNM}$ quantity. Short lines show the magnitude of additional gain 
(negative $E^{NM}-E^{WNM}$ values) or loss (positive $E^{NM}-E^{WNM}$ values) 
in binding energies in the presence of NM. Short-dashed, solid and long-dashed
lines are used for total signature $r=-1$, $r=+1$ and $r=\pm i$ of the configuration,
respectively. The configurations are labeled by the particle (p) and/or hole (h) 
states with respect to the $^{32}$S SD core: the configurations with the holes 
(particles) are shown to the left (right) of $^{32}$S.
}
\label{multi}
\end{figure}

  The nuclei around $^{32}$S in superdeformed minimum are considered 
in the present Section. Their selection is guided in part by the desire 
to compare the CRMF results with the ones obtained in the Skyrme EDF in 
Ref.\ \cite{MDD.00}, where the signature separation induced by time-odd 
mean fields has been found in the excited SD bands of $^{32}$S. The CRMF 
calculations have been performed for some SD configurations in $^{32}$S 
and in neighboring nuclei. The starting point is the doubly magic SD 
configuration $\pi 3^2 \nu 3^2$ in $^{32}$S (further 'SD core') (see Ref.\ 
\cite{Pingst-A30-60}) in which all single-particle orbitals below the 
$N=Z=16$ SD shell gaps are occupied (Fig.\ \ref{routh-s32}). Here
the configurations are labeled by the numbers of occupied proton (p)
and neutron (n) high-$N$ intruder orbitals (the $N=3$ orbitals in our case):
this is commonly accepted shorthand notation $\pi N^n \nu N^p$ of the
configurations in high-spin physics \cite{AKR.96}. Then the configurations 
in the nuclei under consideration 
(Fig.\ \ref{multi}) are created by either adding particles into the 
$[202]5/2^{\pm}$ orbital(s) or/and creating holes in the $[330]1/2^{\pm}$ 
orbitals: these are the orbitals active in signature-separated configurations 
discussed in Ref.\ \cite{MDD.00}.

  Similar to the results shown in Sects.\ \ref{Bind-light} and \ref{Ce-isotopes}, 
the NM leads to additional binding in the configurations of odd mass nuclei (the 
configurations in $^{33}$S and $^{33}$Cl created by adding a particle 
to the SD core or the configurations in $^{31}$P and $^{31}$S created by removing 
a particle from the SD core, see Fig.\ \ref{multi}). This additional binding does 
not depend on the signature of the blocked state.

  Fig.\ \ref{multi} shows that additional binding due to NM is smaller for 
the configurations with blocked proton state as compared with the ones with
blocked neutron state. For example, the configurations in $^{31}$P and $^{31}$S 
are built on the same blocked Nilsson state. However, additional binding due to NM is 
smaller in odd-proton nucleus ($^{31}$P) than in odd-neutron one ($^{31}$S). 
Similar situation also exists in $^{33}$S and $^{33}$Cl. These results are consistent 
with a general systematics (Sect.\ \ref{Mass-dep-en}) which shows that additional 
binding due to NM is smaller in the proton subsystem than in the neutron one. 

 The results of perturbative calculations for the configuration with the proton 
hole in $\pi [330]1/2^-$ in odd-proton $^{31}$P nucleus and for the
configuration with the neutron hole in $\nu [330]1/2^-$ in odd-neutron $^{31}$S 
nucleus are shown in Table \ref{NM-prot}. These hole configurations are formed by 
removing either proton ($^{31}$P) or neutron ($^{31}$S) from the $N=Z$ $^{32}$S 
SD core. One can see that the decrease of additional binding due to NM on going
from neutron to proton configuration of the same structure can be traced to the 
changes in the particle energy $\Delta E_{part}^{pert}$  from $-0.349$ MeV in 
odd-neutron $^{31}$S nucleus to $-0.250$ MeV in odd-proton $^{31}$P. This explains 
the major part of the change in the $\Delta E_{tot}^{pert}$ quantity on going from 
odd-neutron $^{31}$S 
($\Delta E_{tot}^{pert}=-0.165$ MeV) to odd-proton $^{31}$P 
($\Delta E_{tot}^{pert}=-0.100$ MeV). The contributions of other terms 
into $\Delta E_{tot}^{pert}$ on going from odd-neutron 
$^{31}$S to odd-proton $^{31}$P nucleus are smaller: 0.020, 0.001, and 0.013 MeV for 
the $\Delta E_{\omega}^{SL}$, $\Delta E_{\rho}^{SL}$ and $\Delta E_{Coul}$ terms, respectively.

 In perturbative calculations, the changes in particle energy $\Delta E_{part}^{pert}$ can be 
easily related to the energy splitting $\Delta E_{split}$ between the blocked state and its 
unoccupied time-reversal counterpart through $\Delta E_{part}^{pert}\approx - \frac{1}{2} 
\Delta E_{split}$ since the sum over the energies of other occupied single-particle states 
is the same in the calculations with and without NM because the polarization effects are 
absent (Sect.\ \ref{Pol-odd}). The energy splittings between different signatures of the 
blocked $[330]1/2$ state are $\Delta E_{split}=0.653$ MeV and $\Delta E_{split}=0.476$ MeV 
for odd-neutron ($^{31}$S) and odd-proton ($^{31}$P) nuclei, respectively. This result 
clearly indicates that the contributions of the Coulomb force to the proton single-particle 
energies in the presence of NM are at the origin of the fact that additional binding due 
to NM is smaller for odd-proton nuclei as compared with odd-neutron ones. The analysis
of $^{33}$S and $^{33}$Cl leads to the same conclusions.

\begin{table}[h]
\caption{The $\Delta E_{i}^{pert}= (E_i^{NM}-E_i^{WNM})^{pert}$ quantities for different terms 
of the total energy (Eq.\ (\protect\ref{Etot})) for the configurations discussed in the text. 
Only terms affected by NM in perturbative calculations are shown here.}
\label{NM-prot}
\vspace{0.5cm}
\begin{center}
\begin{tabular}{|c|c|c|c|c|} \hline
Quantity            &      $^{31}$S      &  $^{31}$P  & $^{33}$S   & $^{33}$Cl \\ \hline
     1              &       2           &         3  &    4      &      5    \\ \hline
$\Delta E_{part}^{pert}$           &   $-0.349$        &   $-0.250$ & $-0.198$  & $-0.136$  \\
$\Delta E_{\omega}^{SL [pert]}$     &   $-0.168$      &   $-0.148$ & $-0.093$  & $-0.080$  \\
$\Delta E_{\rho}^{SL [pert]}$       &   $-0.016$      &   $-0.015$ & $-0.010$  & $-0.009$  \\
$\Delta E_{Coul}^{pert}$           &   $0.0$          &   $0.013$  &    0      & $0.010$   \\
$\Delta E_{tot}^{pert}$            &  $-0.165 $       &   $-0.100$ & $-0.095$  & $-0.057$  \\ \hline \hline
$\Delta E_{kin}^{pert}$            &  $-0.348 $       &   $-0.276$ & $-0.198$  & $-0.155$  \\ \hline 
\end{tabular}
\end{center}
\label{Table-pert}
\end{table}

\begin{table*}[ht]
\caption{The changes $\Delta E_{i}= E_i^{NM}-E_i^{WNM}$ in different terms of the
total energy (Eq.\ (\protect\ref{Etot})) 
in the SD configurations $A \equiv \pi [202]5/2^- \otimes \nu [202]5/2^- (r=-1)$ and 
$B \equiv \pi [202]5/2^+ \otimes \nu [202]5/2^- (r=+1)$ of $^{34}$Cl induced by NM 
(columns 3-6). Second column shows the absolute energies [in MeV] of different 
energy terms in the case when NM is neglected. The configurations are given with respect of 
doubly magic SD configuration in $^{32}$S. Fully self-consistent (columns 2, 3, and 5) and
perturbative (columns 4 and 6) results are presented.}
\label{Table-Cl34}
\vspace{0.5cm}
\begin{center}
\begin{tabular}{|c|c|c|c|c|c|} \hline
Quantity    &  $E_i^{WNM}(A,B)$  &  $\Delta E_i(A)$ & $\Delta E^{pert}_i(A)$ &
$\Delta E_i(B)$ & $\Delta E_i^{pert}(B)$ \\ \hline
1 & 2 & 3 & 4 & 5 & 6 \\ \hline
$E_{part}$          & $-835.845$   &   $-1.471$ & $-0.791$  & $-0.004$  &  $-0.003$ \\
$E_{\sigma}$        & $-4415.660$    &  $-7.862$ &            & $+0.004$  &            \\
$E_{\sigma NL}$      &  $84.884$    &  $-0.036$ &            & $-0.001$  &            \\
$E_{\omega}^{TL}$    & $3698.240$   &  $7.126$  &            & $-0.003$  &            \\
$E_{\omega}^{SL}$    &   $ 0.0$     &  $-0.414$ & $-0.414$   & 0.0       & $-0.001$   \\
$E_{\rho}^{TL}$      &   $0.061$    &  0.0      &            & 0.0       &            \\
$E_{\rho}^{SL}$      &   $ 0.0$     &  0.0      &            & $-0.043$  & $-0.043$   \\
$E_{Coul}$          &  $59.832$    &  $0.052$  &  0.025     & $-0.002$  & $-0.003$    \\
$E_{cm}$            &  $ -9.492$   &  0.0      &            & 0.0       &            \\
$E_{tot}$           &  $-272.693$  & $-0.376$  & $-0.402$   & 0.041     &  $0.043$   \\ \hline \hline
$E_{kin}$           &  479.210     & $-0.103$  & $-0.841$   & $-0.002$  &  $0.003$   \\ \hline
\end{tabular}
\end{center}
\end{table*}

   The situation is more complicated in odd-odd nuclei ($^{30}$P and $^{34}$Cl) in 
which considerable energy splitting between the $r=+1$ and $r=-1$ configurations is 
obtained in the calculations. The microscopic mechanism of binding modifications is 
illustrated in Table \ref{Table-Cl34} on the example of configurations A and B in 
$^{34}$Cl.

  NM provides additional binding of around 0.4 MeV in the configuration A which has 
signature $r=-1$. In this configuration, proton and neutron currents due to the 
occupation of proton and neutron 5/2[202]$^-$ states are in the same direction 
which results in appreciable total 
baryonic current. This baryonic current leads to sizable modifications in the $E_{part}$, 
$E_{\sigma}$, $E_{\omega}^{TL}$, $E_{\omega}^{SL}$ terms (Table \ref{Table-Cl34}). 
These are precisely the same terms which are strongly affected by NM in odd-mass 
nuclei, see Sect.\ \ref{Pol-odd}. The fermionic contribution into $E^{NM}-E^{WNM}$ 
(the $\Delta E_{part}$ term) is defined by more or less equal contributions from time-odd 
mean fields and the polarization effects in time-even mean fields. On the contrary,  
time-odd mean fields define only $\approx 1/3$ ($\Delta E_{\omega}^{SL}=-0.413$ MeV) of the 
mesonic contribution into $E^{NM}-E^{WNM}$, while the rest is due to the polarization 
effects in time-even mean fields (the $\Delta E_{\sigma}$, $\Delta E_{\omega}^T$ terms). 

  NM leads to the loss of binding in the configuration B 
which has $r=+1$. In this configuration, the proton and neutron currents due 
to the $\pi [202]5/2^+$ and $\nu [202]5/2^-$ states are in opposite directions, so 
the total baryonic current is very close to zero. As a result, the impact of NM is 
close to zero for the majority of the terms in Eq.\ (\ref{Etot}) (see Table 
\ref{Table-Cl34}).  The only exception is the $E_{\rho}^{SL}$ term which represents 
space-like component of the isovector-vector $\bf \rho$-field. This term depends on 
the difference of proton and neutron currents (Eq.\ (\ref{rho-space})), which for 
the present case of opposite currents gives a non-zero result. As follows from Table 
\ref{Table-Cl34}, this term is predominantly responsible for the loss of binding 
due to NM in the configuration B.
  
 It is well known that many physical quantities are additive in the calculations
without pairing (see Ref.\ \cite{MADLN.07} and references therein).  The additivity
principle states that the average value of a one-body operator $\hat{O}$ in a given
many-body configuration $k$, $O(k)$, relative to the average value in the core configuration,
$O^{core}$, is equal to the sum of effective contributions $o^{eff}_{\alpha}$ of particle 
and hole states by which the $k$-th configuration differs from that of the core \cite{MADLN.07}
\begin{eqnarray}
\delta O(k) = O(k)-O^{core}=\sum_{\alpha} c_{\alpha}(k) o^{eff}_{\alpha}
\end{eqnarray}
Coefficients $c_{\alpha}(k)$  ($c_{\alpha}(k) =0,\,\,{\rm or} \,\,+1\,\,{\rm or}\,\, -1$) define
the s.p. content of the configuration $k$ with respect to the core configuration
(see Ref.\ \ \cite{MADLN.07} for details). Let us check whether additional binding due 
to NM (the $\Delta E_{tot}=E^{NM}-E^{WNM}$  quantity) is additive. 
The doubly magic SD configuration $\pi 3^2 \nu 3^2$ in even-even $^{32}$S nucleus is used 
as a core for this analysis: the effective contributions $\delta E_{i}^{eff}$ of particle 
state(s) to $\Delta E_{tot}$ are given by $\delta E_{i}^{eff} = [E_{i} ({\rm nucleus\,\,A}) - 
E_{i} ({\rm core})]^{NM}- [E_{i} ({\rm nucleus\,\, A}) - E_{i} ({\rm core})]^{WNM}
 = E_{i}^{NM} ({\rm nucleus\,\, A})- E_{i}^{WNM}({\rm nucleus\,\, A}) = \Delta E_{i}({\rm nucleus\,\, A}))$ 
because the core configuration  is not affected by NM. Thus, the additivity implies that 
$\Delta E_{tot}(^{34}{\rm Cl}(r=+1)) = \Delta E_{tot}(^{33}{\rm S}) + \Delta E_{tot}(^{33}{\rm Cl})$
($\Delta E_{tot}(^{34}{\rm Cl}(r=-1)) = \Delta E_{tot}(^{33}{\rm S}) - \Delta E_{tot}(^{33}{\rm Cl})$)
for the situation when the proton and neutron currents in $^{34}$Cl are in the same 
(opposite) directions. Fig.\ \ref{multi} clearly shows that additivity conditions are not 
fulfilled and that additional binding due to NM is not additive in self-consistent calculations. 
The analysis involving odd-odd $^{30}$P and odd $^{31}$P, $^{31}$S nuclei leads to the same 
conclusion (see Fig.\ \ref{multi}).

 The additivity is also violated in perturbative calculations: the comparison of 
Tables \ref{Table-pert} (columns 4 and 5) and \ref{Table-Cl34} (columns 4 and 6)
reveals that the conditions 
$\Delta E_{i}^{pert}(^{34}{\rm Cl}(r=\pm1)) = \Delta E_{i}^{pert}(^{33}{\rm S}) 
\pm \Delta E_{i}^{pert}(^{33}{\rm Cl})$ 
are violated both for the total energy ($i=tot$) and individual components of the total 
energy $(i=part,\,\, _{\omega}^{SL},\,\, _{\rho}^{SL},\,\, Coul)$. The analysis of 
$\Delta E_{part}^{pert}$ (this terms provides the largest contribution to $\Delta E_{tot}^{pert}$
(see Tables \ref{Table-pert} and \ref{Table-Cl34})) allows to understand the origin of the 
violation of additivity for the $\Delta E_{tot}^{pert}$ quantity. 
In odd-proton $^{31}$Cl nucleus, 
$\Delta E_{part}^{p [pert]} \approx - 1/2 \Delta E_{split}^{p}$ ($\Delta E_{split}^{p}$ is 
the energy splitting between blocked proton state and its signature counterpart)
and $\Delta E_{split}^{p}$ depends predominantly on the proton current induced by odd 
proton. The same is true in odd-neutron $^{33}$S nucleus where $\Delta E_{split}^{n}$ 
depends predominantly on the neutron current induced  by odd neutron. 
Additivity principle implies $\Delta E_{part}^{odd-odd [pert]} \approx - (1/2 \Delta E_{split}^{p} 
+ 1/2 \Delta E_{split}^{n})$ for the $^{34}{\rm Cl}(r=+1)$ configuration, in which the proton 
and neutron currents are in the same direction. However, proton $\Delta E_{split}^{p [odd-odd]}$ 
(neutron $\Delta E_{split}^{n [odd-odd]}$) energy splitting between blocked proton (neutron) 
state and its time-reversal counterpart in odd-odd nuclei depend on  total baryonic 
(proton+neutron) current in this configuration.  On the contrary, the additivity principle 
implies that these proton and neutron quantities depend on individual proton and neutron 
currents in odd-odd nucleus, respectively.  This total current is approximately two times 
stronger than individual (proton or neutron) currents in odd mass nuclei. As a consequence, 
$\Delta E_{split}^{p [odd-odd]}$ and $\Delta E_{split}^{n [odd-odd]}$ values in odd-odd mass 
nucleus are larger than the same quantities ($\Delta E_{split}^{p}$, $\Delta E_{split}^{n}$)
in odd-mass nuclei by a factor close to 2.  As a result, $\Delta E_{part}^{pert}(^{34}{\rm Cl}(r=+1)) 
\approx 2 (\Delta E_{part}^{pert}(^{33}{\rm S}) + \Delta E_{part}^{pert}(^{33}{\rm Cl})$ 
(see Tables \ref{Table-pert} and \ref{Table-Cl34}) which clearly indicates the violation
of additivity for the $\Delta E_{part}^{pert}$ quantity (and for the $\Delta E_{tot}^{pert}$ 
quantity).

\begin{figure}
\vspace{0.5cm}
\centering
\includegraphics[width=8.0cm]{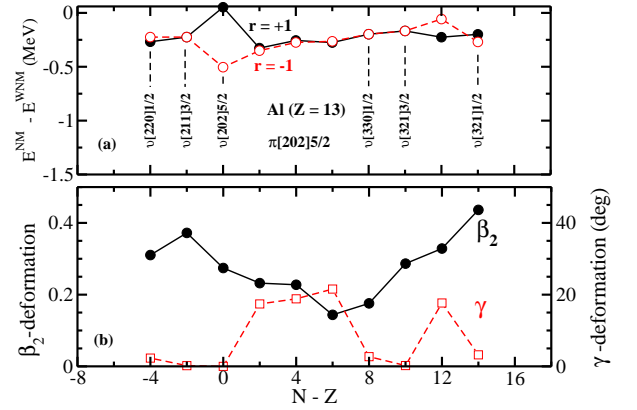}
\caption{(Color online) The impact of NM on binding energies of the lowest configurations 
in odd-odd Al nuclei. Upper panel shows the $E^{NM}-E^{WNM}$ quantity for different signatures. 
The structure of the blocked states are shown by the Nilsson labels only in the cases when 
the configurations are near-prolate. The same state is blocked in the proton 
subsystem of all nuclei. The bottom panel shows the $\beta_2$ and $\gamma$-deformations of 
the configurations under study.}
\label{odd-odd-al}
\end{figure}

\begin{figure}
\vspace{0.5cm}
\centering
\includegraphics[width=8.0cm]{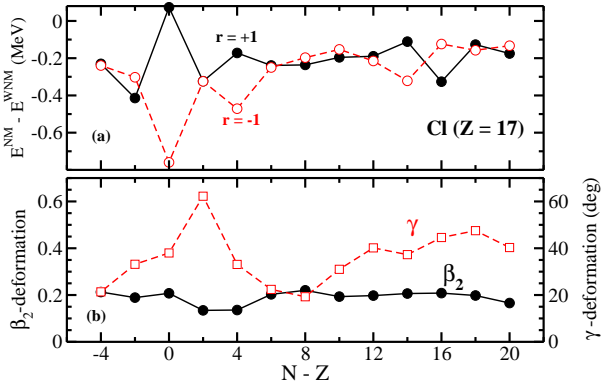}
\caption{(Color online) The same as in Fig.\ \ref{odd-odd-al} but for the lowest 
configurations in odd-odd Cl nuclei.}
\label{odd-odd-cl}
\end{figure}

\begin{figure}
\vspace{0.8cm}
\centering
\includegraphics[width=8.0cm]{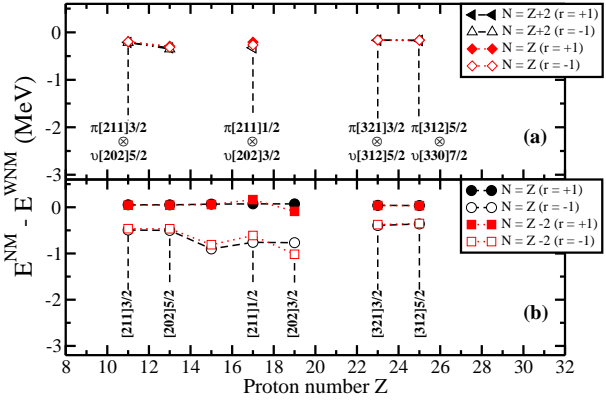}
\caption{(Color online) The impact of NM on binding energies of the configurations
in odd-odd nuclei in the vicinity of the $N=Z$ line as a function of proton 
number $Z$.  Panel (a) shows the results for the configurations with 
different blocked proton and neutron states, while panel (b) shows the results for the 
configurations with the same blocked proton and neutron states. The structure of the blocked 
states are shown by the Nilsson labels only in the cases when the configurations 
are near-prolate. Note that in panel (b) only one Nilsson label is shown since
blocked proton and neutron states have the same structure. The results are shown only in 
the cases when the convergence has been achieved for both $N=Z$ and $N=Z-2$ (or 
$N=Z+2$) nuclei.}
\label{nznz}
\end{figure}

  Fig.\ \ref{multi} also shows the results for the 4-particle excited SD states 
$\pi ({\rm ab})\otimes \nu ({\rm ab})$ in $^{32}$S, for which the calculated rotational 
structures display the signature separation induced by time-odd mean fields 
\cite{MDD.00,Pingst-A30-60}. The configurations are formed by exciting proton 
and neutron from the $[330]1/2^{-}$ orbitals below the $N=16$ and $Z=16$ SD shell 
gaps into the $[202]5/2^{\pm}$ orbitals located above these gaps. They have the 
$\pi 3^1 \nu 3^1$ structure in terms of intruder orbitals. When NM is neglected 
these four configurations are degenerated in energy. This degeneracy is broken 
and additional binding, which depends on the total signature of the configuration  
(0.907 MeV for the $r=+1$ configurations and 0.468 MeV for the $r=-1$ configurations 
in the calculations with the NL3 parametrization),  is obtained when NM is taken 
into account. The NL1 and NLSH parametrizations of the RMF Lagrangian give very 
similar values of additional binding due to NM. The essential difference between 
the relativistic and non-relativistic calculations lies in (i) the size of the 
energy gap between the $r=+1$ and $r=-1$ configurations and (ii) the impact of 
time-odd mean fields on the energy of the $r=-1$ states. This energy gap is about 
2 MeV in the Skyrme EDF calculations with the SLy4 force \cite{MDD.00}, while it is 
much smaller being around 0.45 MeV in the CRMF calculations with the NL1, NL3 and 
NLSH parametrizations. The energies of the $r=-1$ states are not affected by time-odd
mean fields in the Skyrme EDF calculations \cite{MDD.00}, while appreciable additional 
binding is generated by NM for these states in the CRMF calculations (Fig.\ 
\ref{multi}).
 
  Figs.\ \ref{odd-odd-al} and \ref{odd-odd-cl} show the results of the calculations 
for ground state configurations in odd-odd Al and Cl nuclei. The calculations 
suggest that signature separation due to time-odd mean fields is also expected in the 
configurations of odd-odd nuclei located at zero or low excitation energies. 
The signature separation is especially pronounced in the $N=Z$ $^{26}$Al (the 
$\pi [202]5/2 \otimes \nu [202]5/2$
configuration) and $^{34}$Cl nuclei. This is because proton and neutron currents in 
these configurations are almost the same both in strength and in spatial distribution. 
As a result, their contribution to the total energy is large when these currents 
are in the same direction (the $r=-1$ configurations) and close to zero when these 
currents are in opposite directions (the $r=+1$ configurations). Note that $^{26}$Al 
is axially deformed while $^{34}$Cl is triaxially deformed with 
$\gamma \sim 30^{\circ}$. However, both of them show the enhancement of the signature 
separation at $N=Z$.

  The signature separation is rather small for the majority of nuclei away from the $N=Z$ 
line. This is a consequence of the fact that the strength of the currents in one 
subsystem  (and thus the impact of NM on binding energies) is much stronger than in 
another subsystem. As a result, there is no big difference (large signature separation) 
between the cases in which proton and neutron currents are in the same and opposite 
directions. However, some nuclei away from the $N=Z$ line also show appreciable signature 
separation. These are $^{38}$Al and $^{38,48,50}$Cl nuclei (Figs.\ \ref{odd-odd-al},
\ref{odd-odd-cl}) for which the strengths of proton and neutron currents (but not necessarily 
the spatial distribution of the currents) are of the same order of magnitude.

  It was suggested in Ref.\ \cite{S.99} that the effects of time-odd mean fields are enhanced 
at the $N=Z$ line.  However, Fig.\ \ref{nznz} clearly shows that the enhancement of signature 
separation  is not restricted to the $N=Z$ line. Indeed, signature separation of the 
configurations based on the same combination of blocked proton and neutron states
are very similar in the $N=Z$ and $N=Z\pm 2$ nuclei despite the 
fact that the deformations of compared nuclei differ  sometimes appreciably. There are considerable 
signature separation in the configurations based on the same blocked proton and neutron states 
in the $N=Z$ and $N=Z-2$ nuclei (Fig.\ \ref{nznz}b). On the other hand, almost no signature 
splitting is observed in the $N=Z$ and $N=Z+2$ nuclei when the configurations are based 
on different blocked proton and neutron states (Fig.\ \ref{nznz}a). This suggests that the enhancement 
of signature splitting is due to similar proton and neutron current distributions (see discussion 
in previous paragraph).

  When considering odd-odd nuclei one has to keep in mind that the present approach 
takes into account only the part of correlations between blocked proton and neutron 
and neglects the pairing. In particular, the residual interaction of unpaired proton 
and neutron leading to the Gallagher-Moshkowski doublets of two-quasiparticle states 
with $K_>=\Omega_p+\Omega_n$ and $K_<=|\Omega_p-\Omega_n|$ \cite{GM.58,BPO.76} is not 
taken into account. Thus, future development of the model is required in order to compare 
directly the experimental data on odd-odd nuclei with calculations. This question will 
be discussed in more details in a forthcoming manuscript \cite{AA.09-2}.

\section{Conclusions}
\label{Sect-concl}

  Time-odd mean fields (nuclear magnetism) have been studied at no rotation in a 
systematic way within the framework of covariant density functional theory by performing 
the blocking of the single-particle states with fixed signature. The main results can be 
summarized as follows:

\begin{itemize}

\item
   In odd-mass nuclei, nuclear magnetism always leads to an additional binding 
indicating its attractive nature in the CDFT. This additional binding only weakly 
depends on the parametrization of the RMF Lagrangian. On the contrary, time-odd 
mean fields in Skyrme EDF can be attractive and repulsive and show considerable 
dependence on the parametrization of density functional. This additional binding 
is larger in odd-neutron states than in odd-proton ones in the CDFT framework.
The underlying microscopic mechanism of additional binding due to NM has been
studied in detail. The perturbative results clearly indicate that additional 
binding due NM is defined mainly by time-odd fields and that the polarization 
effects in fermionic and mesonic sectors of the model cancel each other to a 
large degree.

\item
  Additional binding due to NM can have a profound effect on the properties of 
odd-proton nuclei in the ground and excited states in the vicinity of the proton-drip 
line. In some cases it can transform the nucleus which is proton unbound (in the
calculations without NM) into the nucleus which is proton bound. This additional
binding can significantly affect  the decay properties of proton unbound nuclei
by (i) increasing the half-lives of proton emitters (by many orders of magnitude in 
light nuclei) or (ii) moving the $Q_p$ value inside or outside the $Q_p$ window 
favorable for experimental observation of proton emission.

\item
 Relative energies of different (quasi)particle states in medium and heavy
mass nuclei are only weakly affected by time-odd mean fields. This is because 
additional bindings due to NM show little dependence on blocked single-particle 
state. As a result, the present investigation suggests that time-odd mean 
fields can be neglected in the fits of covariant density functionals aimed 
at accurate description of the energies of the single-particle  states.

\item 
  The phenomenon of signature separation \cite{MDD.00} and its microscopic mechanism 
have been investigated in detail. It was shown that this phenomenon is active also
in the configurations of odd-odd nuclei. It is enhanced for configurations 
having the same blocked proton and neutron states; this takes place either at 
ground state or at low excitation energy in the nuclei at or close to the $N=Z$ line. 
Some configurations away from the $N=Z$ line also show this effect but signature 
separation is appreciably smaller.

\end{itemize}

  The present investigation has been focused on the study of time-odd mean fields in 
the CDFT with non-linear parametrizations of the Lagrangian. Point coupling \cite{BMMR.02} 
and density dependent meson-nucleon coupling \cite{LNVR.05} models are other classes of 
the CDF theories. It is important to compare them in order to make significant progress 
toward a better understanding of time-odd mean fields. The work in this direction 
is in progress, and the results will be presented in a forthcoming manuscript 
\cite{AA.09-2}.

\section{Acknowledgments}

 The material is based upon work supported by the Department of Energy 
under grant Number DE-FG02-07ER41459. Private communication of P.\ Ring 
\cite{R-priv-09} is greatly appreciated.

\end{document}